\documentstyle{elsart}
\newcommand{\be}{\begin{equation}\label}
\newcommand{\ee}{\end{equation}}

\begin{document}
\input{epsf.sty}
\begin{frontmatter}
\title{
Localization and Coherence 
in Nonintegrable Systems
}
\author{Benno Rumpf}
\address{Max-Planck-Institut f\"ur Physik komplexer Systeme
N\"othnitzer Stra\ss e 38, 01187 Dresden, Germany, 
email: benno@mpipks-dresden.mpg.de}
\author{Alan C. Newell}
\address{Mathematics Department, University of Arizona, 
617 North Santa Rita, 
Tucson, Arizona 85721, USA 
}

\begin{abstract}

We study the irreversible dynamics of 
nonlinear, nonintegrable Hamiltonian oscillator chains approaching  
their statistical asymptotic states. 
In systems constrained by more than one conserved 
quantity, the partitioning of the conserved quantities 
leads naturally to localized 
and coherent structures. If the phase space is compact, the 
final equilibrium state is governed by entropy maximization and the coherent 
structures are stable lumps. 
In systems where the phase space is 
not compact, the coherent structures can be collapses represented in 
phase space by a heteroclinic connection of some unstable saddle to 
infinity. 
\end{abstract}
\end{frontmatter}

keywords: Localized structures, statistical physics
\section{Introduction}
The formation of coherent structures by a continuing self focusing 
process is a widespread phenomenon in dispersive nonlinear wave systems. 
As a result 
of this process, high peaks of some physical field emerge from 
a low amplitude noisy background. 
For optical waves in Kerr-nonlinear media this results in a  
self-enhanced increase of light intensity in a small sector of a laser beam 
while the intensity in the neighborhood of this bright spot decreases 
\cite{kel}. 
Related phenomena occur in diverse systems from hydrodynamics 
to collapsing Langmuir waves in a plasma \cite{zcol} and in 
numerical algorithms for partial differential equations 
\cite{new}, \cite{bns} . 
\\
A common feature of the wave dynamics of these systems is the 
comparable strength of the dispersion and the nonlinearity. 
But, self focusing phenomena are radically different in integrable and 
nonintegrable systems. In integrable systems \cite{zint}, the 
peaks appear and disappear in a quasiperiodic manner reflecting the 
phase space structure of nested tori. This behavior is usually encountered 
on time scales that are short enough so that generic nonintegrable 
contributions to the dynamics may be neglected. 
Significant changes occur on time scales where the nonintegrability 
is relevant \cite{zak},\cite{dya}. The solution's shape becomes more irregular 
\cite{abhe},\cite{mojo}, \cite{mls},\cite{cal} 
and the periodic breathing of the peaks turns into a more persistent 
state. These peaks can merge into stronger ones while radiating 
low-amplitude waves. The irreversible character of the system 
becomes apparent and its behavior is driven by 
statistical mechanics \cite{jor},\cite{ras},\cite{rune} 
as the solution's trajectory tries to explore 
more of the available phase space. As it does this, it must, at the same 
time, respect the preservation of conserved quantities. 
The phase space shell of constant conserved quantities determines 
the system's most favorable macrostate and the resulting 
dynamics can lead to a local gathering of 
the amplitude while the system explores the phase space shell. 
We will see 
in situations with more than one conserved quantity that the spatial 
structure of the state which is statistically preferred now contains 
coherent peaks. If the phase space is compact, these peaks can be 
local , time independent and stable. If the phase space is not compact, 
the peaks can be collapsing filaments which produce singularities. 
\\
In this paper we study three systems which 
typify the focusing behavior observed in nonintegrable Hamiltonian systems 
with more than one conserved quantity,  
namely the Landau-Lifshitz equation for a classical Heisenberg 
spin chain, various versions of the discrete nonlinear Schr\"odinger equation,
and a leapfrog discretization of the Korteweg-de Vries equation. 
Various modifications of these systems help to identify the role 
of the conserved quantities in the formation of coherent structures. 
The spatial discreteness of all these systems avoids 
fluctuations on infinitesimal scales. \\
The first case study investigates 
the Landau-Lifshitz equation for the classical 
spin chain in one spatial dimension. We study the 
long time behavior of those configurations in 
which most spins are close to the north pole. The two conserved 
quantities are the energy and the magnetic moment. 
Almost constant states undergo a series of modulational instabilities 
and the system begins to oscillate as if it were integrable. 
Nonintegrability, however, leads to non recurrence as localized peaks appear 
which merge from time to time leading to even larger peaks and radiating 
some energy. 
The phase space is compact, and 
one can 
compute to good accuracy the thermodynamic potentials of the 
system by separating 
the low-amplitude spin waves and the strongly nonlinear components. 
Depending on the initial value of the energy and the magnetization, 
entropy maximization leads to a state where some of the 
magnetization must be put into local structures bounded by domain 
walls for which the spin of each contained lattice point is close 
to the south pole. 
Numerical simulations very clearly support our simple analytical predictions 
which are based on thermodynamic considerations. 
\\
A close relative of the Heisenberg spin chain is found 
by taking the small amplitude 
limit where all spins are close to the north pole. 
Deviations are described by the focusing nonlinear Schr\"odinger 
equation. A similar dynamics is observed. In this study, we can also 
investigate the effects of exact integrability by using the Ablowitz-Ladik 
algorithm \cite{abla}. In these simulations, we find no irreversible behavior, 
no peak fusion, no relaxation, only quasi-periodic behavior. 
Likewise, we get qualitatively different results if we take a model 
which breaks the rotational symmetry and because of this the particle 
number is no longer conserved. 
As a consequence it is not necessary for the system to 
develop coherent structures in order to maximize its entropy as it no longer 
has to be concerned about the second constant of motion when its trajectory 
explores the accessible phase space. \\
The last case study concerns the leapfrog algorithm for the numerical 
integration of the Korteweg-deVries equation. 
It was observed in previous works \cite{new},\cite{bns} that the 
leapfrog algorithm for the Korteweg-deVries equation always 
develops singularities. 
After (usually) a very long time, the amplitudes in some local neighborhood 
rapidly diverges. We 
demonstrate that this collapse again takes 
on an organized coherent form. \\
How and why does the system develop such local objects? The reason is 
again statistical. At an early stage, one can again observe the gathering  
of one of the conserved quantities in coherent structures. 
The system's phase space is not compact, however, so that a strong 
nonequilibrium process prevails finally. 
We find a rapidly growing localized 'monster'-solution 
(so called because of its likeness in shape to the Loch Ness monster) that 
has a canonical structure. This solution can originate from coherent 
structures or, most frequently, from a long wave instability of the 
low-amplitude noisy background. We discuss the similarity of this 
process to the collapse behavior of the 
two dimensional focusing nonlinear Schr\"odinger equation, 
where the conservation laws necessitate a net inverse particle flux 
to small wavenumbers. \\
This paper is arranged as follows: 
In section 2, we present the Landau-Lifshitz equation, the discrete nonlinear 
Schr\"odinger equation and the leapfrog discretization of the 
Korteweg - de Vries equation and discuss some of their properties. 
In section 3 we present numerical studies of these equations. 
In particular, we study thermodynamic quantities 
during the focusing process. Their significance is also 
demonstrated by 
discussing modified equations with either more or with fewer conserved 
quantities as well as an equation of defocusing type. 
In section 4 we will give a statistical interpretation of the numerical 
findings. 
By computing the thermodynamic potentials of the spin-system, we find 
the connections between global features of the pattern and the 
conserved quantities. 
Macroscopic properties of the final state are computed. The 
discretized KdV equation has no such state of thermal equilibrium. 
We identify the rapid divergence of the amplitudes with the exploration 
of the non-compact phase space shell and we suggest that there is much 
similarity between this behavior and the condensation and collapse behavior 
seen in the focusing nonlinear Schr\"odinger equation. \\
Figures of similar contents are grouped together and their order  
sometimes deviates from the sequence of their references in the text.
\section{Nonintegrable systems with constraints}
\subsection{Time-continuous systems}
\subsubsection{The Landau-Lifshitz equation}
The anisotropic Heisenberg spin chain is particularly suitable 
for the study of self-focusing phenomena. 
This system contains the 
generic properties of equations of nonlinear Schr\"odinger type that 
lead to self-focusing and it is easy to 
investigate from the statistical point of view. 
The Landau-Lifshitz equation \cite{koiv}  
\begin{equation}\label{spinchain}
\dot{\bf S}_n={\bf S}_{n}\times (J({\bf S}_{n-1}+{\bf S}_{n+1})
+ S_{nz}{\bf e}_z)
\end{equation}
is a classical approximation of the dynamics of magnetic 
moments ${\bf S}_n=(S_{xn},S_{yn},S_{zn})$ at lattice sites $n$. 
$\dot{\bf S}_n$ is perpendicular to ${\bf S}_n$. Therefore the moduli 
of the spin vectors are conserved and one may set $|{\bf S}_n|=1$. 
The phase space of a chain of $N$ spins is a product of $N$ such spheres. 
$S_z$ the component of ${\bf S}$ along the rotational symmetry axis. 
The northern and southern hemispheres are equivalent since 
(\ref{spinchain}) is invariant under the transformation 
$(S_{xn},S_{yn},S_{zn})\rightarrow (-S_{xn},S_{yn},-S_{zn})$. 
\\
There are two trivial homogeneous equilibrium states where all the spins 
point either to the north pole $S_z=1$ or to the south pole $S_z=-1$. 
Throughout this paper we only consider solutions where most of the spins 
are close to the north pole. 
\subsubsection{The discrete nonlinear Schr\"odinger equation}
The long-wavelength dynamics of spins which deviate slightly 
from the north pole $S_z= 1$ 
is given by the focusing discrete nonlinear Schr\"odinger (DNLS) equation  
\begin{equation}\label{nlseq}
i\dot\phi_n=J(\phi_{n+1}+\phi_{n-1}-2\phi_n)+2|\phi_n|^2\phi_n
\end{equation} 
for small values of the complex amplitude $\phi=(S_x+iS_y)/(1+S_z)$. 
The spin chain may thus be regarded as a DNLS which is modified by 
higher order terms. The north pole is corresponds to $\phi=0$ while 
the south pole corresponds to an infinite amplitude. 
\subsubsection{Integrals of motion}
The spin chain and the DNLS equation each have two conserved 
quantities: 
\begin{enumerate}
\item
The Hamiltonian of the DNLS equation 
${\cal H}=\sum_n J(2\phi_n\phi_n^*-\phi_n\phi_{n+1}^*-\phi_n^*\phi_{n+1})-|\phi_n|^4$ is again 
obtained 
as the lowest order of the Hamiltonian of the spin chain 
${\cal H}={\cal H}_J+{\cal H}_a=
\sum_n J(1-{\bf S}_n {\bf S}_{n+1})+(1-S_{zn}^2)/2$. 
The first contribution is a Heisenberg exchange coupling which is minimal  
for homogeneous solutions. The second part is an anisotropic energy 
that has minima at the poles $S_z=\pm 1$ and is maximal at the equator 
$S_z=0$. The stationary spin-up or spin-down solutions are the absolute 
energy minima. 
Fluctuations about these ground states give energy contributions per 
lattice site of the order of $|S_x+iS_y|^2$. 
Similarly, small fluctuations near by the equilibrium state 
$\phi=0$ of the DNLS contribute a coupling energy proportional to 
$|\phi|^2$. 
In contrast to the south pole state of the spin system, 
the energy of a solution $\phi\rightarrow\infty$ 
in the DNLS goes to minus infinity as $-|\phi|^4$. \\
\item
The second conserved quantity of each system, the total magnetization 
${\cal M}=\sum_n S_{zn}$  of the spin chain 
and the modulus-square norm 
('particle number') $\sum_n |\phi_n|^2$ of the DNLS, 
are related to the system's rotational symmetry. 
The superposition of the Hamiltonian and this integral of motion yields a 
Hamiltonian in a rotating frame system. The negative magnetization 
$N-{\cal M}=\sum (1-S_{zn})\approx \sum |S_{xn}+iS_{yn}|^2/2$ 
corresponds to the 
particle number of the DNLS in the lowest order in amplitude. 
This 'particle number' of the spin chain is zero 
for the north pole solution 
and it is two per lattice site for the south pole solution. 
In the DNLS, the particle number diverges for the state 
$|\phi|\rightarrow \infty$. 
\end{enumerate}
In order to contrast the generic behavior of such systems with those of 
(a) integrable systems and (b) systems not constrained by a 
second conservation 
law, we also consider two modified equations of motion. 
The first is the integrable Ablowitz-Ladik discretization of the 
one-dimensional nonlinear Schr\"odinger equation \cite{abla}). 
The second is an 
equation where the second integral is destroyed by a symmetry breaking 
field. \\
Low energetic solutions just above the ground states can be characterized by 
the ratio of the two integrals of motion, i.e. the energy per particle. 
This reveals a major difference between the spin chain and the DNLS. 
For fluctuations near the north pole or near $\phi=0$, the particle 
number is of the order of the energy so that this ratio is of order one 
for both systems. 
Spin-fluctuations near the south pole have the same energy but the  
second conserved quantity ('particle number') is much higher 
so that the energy per particle is proportional to 
$|S_x+iS_y|^2$ and much less than $1$. 
Thus the spin chain has two states with low energies per lattice 
site, one ($S_z\approx 1$) 
with a higher energy per particle  and one ($S_z\approx -1$) with a  
low positive energy per particle. 
This well-defined condensate state of low energy and high particle density 
is a mayor advantage of the spin chain. \\
In contrast, for infinitely high amplitude solutions 
of the DNLS that correspond to $S_z\approx -1$ solutions of the spin chain 
both the energy and the particle density go to infinity. 
The energy per particle diverges proportional to  
$-|\phi|^2$. \\
\subsection{Time-discretized equation of motion}
\subsubsection{The leapfrog-discretization of the Korteweg-de Vries  
equation}
The system of finite difference equations
\begin{equation}\label{kdvdisc}
\begin{array}{rcl}
u_{m+1}(n)&=&v_{m}(n)\\
&&+\tau (u_m(n+2)-2u_m(n+1)+2u_m(n-1)-u_m(n-2)\\
&&-2(u_m(n+1)+u_m(n)+u_m(n-1))(u_m(n+1)-u_m(n-1)))\\
v_{m+1}(n)&=&u_m(n)
\end{array}
\end{equation}
is a leapfrog-type discretization in space and time 
of the completely integrable Korteweg-de Vries (KdV) equation 
$\dot{u}=u_{xxx}-6uu_x$ 
for the real amplitude $u(x,t)$. 
The leapfrog discretization is characterized by a central difference 
$\partial u/\partial t \rightarrow (u_{m+1}-u_{m-1})$. 
The factor $\tau$ of the spatial derivative 
is the time step size. 
The precursor $u_{m-1}$ is identified with the additional variable 
$v_m$ on the right side of the first equation. 
This scheme allows the  
simulation of the partial differential equation 
avoiding the amplitude dissipation that occurs in methods with 
numerical viscosity. 
The term $u_m(n+2)-2u_m(n+1)+2u_m(n-1)-u_m(n-2)$ is the standard 
discretization of $u_{xxx}$. 
The discretization of $uu_x$ was first suggested by Zabusky and 
Kruskal \cite{zakr}. 
It involves a central difference in space 
$u_m(n+1)-u_m(n-1)$ and replaces $u$ by the average 
$(u_m(n+1)+u_m(n)+u_m(n-1))/3$. \\
This discretization suppresses fast-acting nonlinear instabilities. 
Discretizations that do not retain some of the original 
conservation laws lead to fast acting instabilities, since 
single modes diverge rapidly. 
For instance, the mode with the wavenumber $k=2\pi/3$ is driven by 
the nonlinear part of conventional discretizations of the KdV equation.
In contrast, this mode is an exact solution of (\ref{kdvdisc}). 
Linear instabilities of the zero-solution can 
be avoided by a sufficiently small step-size 
$\tau<2/(3\sqrt{3})$. 
\subsubsection{Integrals of motion}
The special feature of the spatial discretization (\ref{kdvdisc}) 
is that it preserves some of the original conserved quantities: 
\begin{enumerate}
\item
$<uv>=\sum_n u_m(n)v_m(n) = const.$ 
corresponds to the conserved quantity $\int u^2 dx$ ('energy' for 
shallow water waves) 
in the original KdV equation. The modulus-square norm 
$<u^2+v^2>=\sum_n u_m(n)^2+v_m(n)^2$ is not conserved. 
\item
$<u>=\sum_n u_{2m}(n)=\sum_{n}v_{2m+1}(n)$ 
and $<v>=\sum_n v_{2m}(n)=\sum_{n}u_{2m+1}(n)$
correspond to $\int u dx$ ('mass' for shallow water waves). 
\end{enumerate}

\section{Numerical studies}
We examine the formation of coherent structures numerically 
in various versions of the spin chain and the DNLS equation as 
well as the leapfrog integration scheme for the KdV equation. 
A typical scenario for the spin chain suggests that the final state 
mainly depends on the amount of the two conserved quantities provided 
by the initial conditions. Simulations of various modifications 
of the DNLS with either more of less integrals of motion clarify 
some  more general conditions for this behavior. 
The simulations of the differential equations apply an Adams routine 
to a chain of 512 (and occasionally 4096) oscillators 
with periodic boundary conditions. 
\subsection{Dynamics of the spin chain}
\subsubsection{Benjamin-Feir instability and reversible dynamics}
\begin{figure}[htb]
\epsfbox{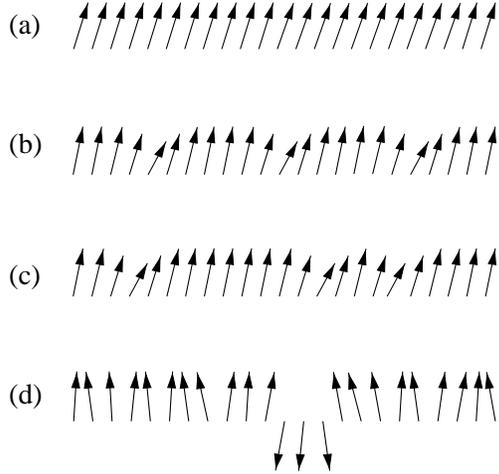}
\caption{
Sketch of the spins for (a) the spatially homogeneous 
solution, (b) the spatially periodic solution resulting 
from a Benjamin-Feir instability, (c) the humps moving towards 
each other merging into compound peaks of spins pointing down (d). 
}
\label{spinsketch}
\end{figure}

Plane wave solutions of the nonlinear Schr\"odinger equation 
are Benjamin-Feir unstable so that 
self-focusing is initiated by long wavelength modulations. 
Similarly, spin-wave solutions of the Heisenberg spin chain 
are unstable under long wavelength perturbations. For instance, 
the homogeneously magnetized solution ${\bf S}_n={\bf S}$ 
(Fig.\ref{spinsketch}a) 
that precesses 
about the symmetry axis ${\bf e}_z$ with the frequency 
$\omega=S_z$ is most unstable under perturbations 
with the wavenumber $k=\sqrt{1-S_z^2}/J$. \\
As a result of this instability, 
small perturbations lead to spatially periodic humps of 
spins approaching the equator while most of the spins come 
closer to the north pole (Fig.\ref{spinsketch}b). 
The trajectory is close to a homoclinic orbit so that the solution returns to 
the almost homogeneous state after reaching the maximum amplitude. 
\begin{figure}[htb]
\epsfbox{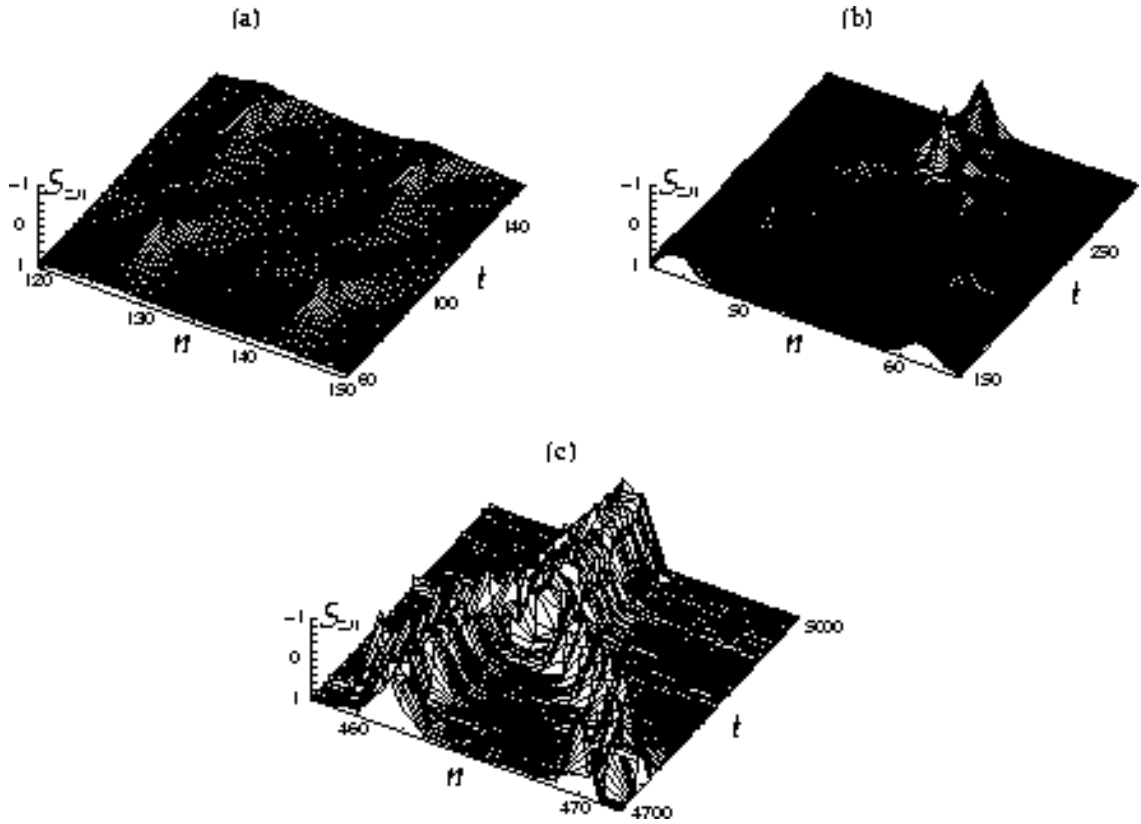}
\caption{$S_{zn}$ as a function of the lattice site $n$ and time. 
(a) portrays the sector ($\alpha$) of Fig.\ref{smax}a, 
(b) is the sector ($\beta$); 
(c) is at ($\gamma$) of Fig.\ref{smax}b. 
}
\label{sprofile}
\end{figure}
\begin{figure}[htb]
\epsfbox{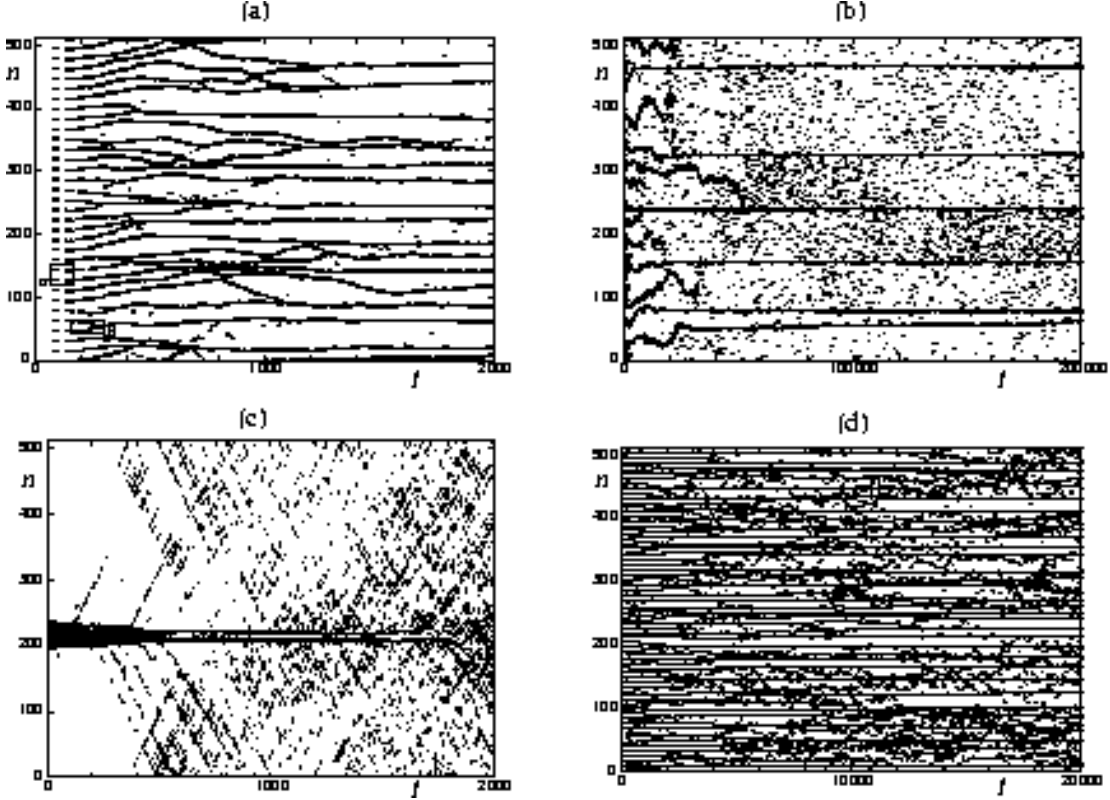}
\caption{Integration of 512 spins $(J=0.4)$ with periodic boundary conditions; 
lattice sites where the spins deviate significantly from the north 
pole ($S_z<0.8$) are marked with dots. Integration 
over (a) 2000 and (b) 200,000 time steps with weakly perturbed homogeneous 
initial conditions $(S_z\approx \sqrt{0.84})$  
(the profile of the z-magnetization at ($\alpha$, $\beta$, $\gamma$) 
are given in Fig.\ref{sprofile}); 
noisy short wave ($k=\pi$) inital conditions with a xenocryst of spins 
deviating strongly from the north pole over 2000 time steps (c); 
defocusing equation 
(negative sign of the potential), weak coupling ($J=0.1$) and 
noisy short wave ($k=\pi$) initial conditions over 20,000 time steps (d). 
}
\label{smax}
\end{figure}
Fig.\ref{sprofile}a shows the profile 
of the z-magnetization in space and time for this solution which is  
almost periodic in space and time. 
Fig.\ref{smax}a shows the pattern of peaks 
(the sites of the spins that differ most 
from the north pole) as a function of time; Fig.\ref{sprofile}a is 
related to box ($\alpha$) of Fig.\ref{smax}a. 
Equidistant humps emerge and disappear periodically in time for $t<200$. 
\subsubsection{Merging of peaks and irreversible dynamics}
The spatially periodic solution arising from the Benjamin-Feir 
instability is itself phase-unstable. 
As a result, the periodic pattern with the wavelength of the initial 
periodic mode becomes modulated on an even larger length scale 
so that the the gaps between the initially equidistant humps start to vary 
(Fig.\ref{spinsketch}c). 
Fig.\ref{smax}a shows 
tiny variations of the distances between the humps 
as the humps start to move at $t\approx60$ . 
\\
The most important phenomenon following the phase instabilities 
is the formation of coherent structures through mergings of peaks. 
Neighboring humps approaching each other 
finally merge into single peaks radiating small fluctuations. 
Fig.\ref{sprofile}b shows the profile of the 
magnetization during the fusion of humps of box ($\beta$) 
in Fig.\ref{smax}a. 
The original periodic solution is smooth, but 
the compound peak resulting from the merging has an 
irregular shape involving huge gradients both in space and in time. 
Its amplitude oscillates irregularly in time, but unlike the original periodic 
solution, it does not vanish any more completely. 
Even those humps that are situated remotely from the first merging processes 
become more persistent in time immediately so that they are traced 
by continuous lines in Fig.\ref{smax}a. \\
Subsequently, more humps fusing into compound peaks increase the average 
distance between neighboring peaks. 
The resulting compound peaks again merge with primary humps 
and with other compound peaks forming even stronger peaks  
(see point ($\gamma$) in Fig.\ref{smax}b
with the magnetization profile of Fig.\ref{sprofile}c). 
\subsubsection{Final equilibrium state}
The increasingly high amplitude of the compound-peaks enables some 
spins to overcome the energy barrier 
of the equator and to flip to the southern hemisphere 
(Fig.\ref{spinsketch}d). 
(Fig.\ref{smax}b) shows that after $2\times 10^5$ time steps 
all peaks have merged into six down-magnetized domains that each 
consist of two or three lattice points. These peaks with 
$S_z\approx-1$ are embedded in a disordered state where the spins deviate 
only slightly from the north pole $S_z=1$. 
The final state is a two domain pattern where most 
of the spins are accumulated in huge up-magnetized domains while 
a few spins condense to small down-magnetized domains. 
The domain of the spin-wave fluctuations near to 
the north pole in the final state remains persistent even 
if the spin-down xenocrysts are removed artificially by flipping 
the down-spins up as $S_z\rightarrow |S_z|$. 
\subsubsection{Transfer of energy}
\begin{figure}[htb]
\epsfbox{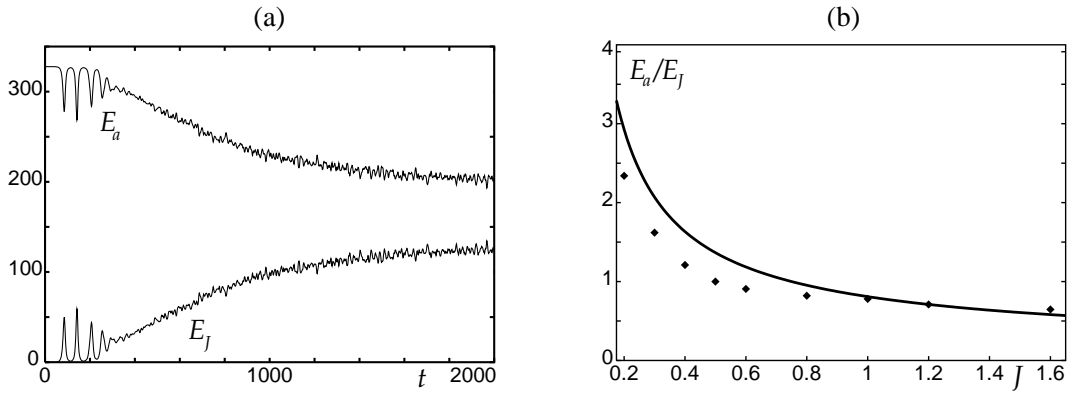}
\caption{(a): The evolution of the anisotropic energy $E_a$ and the coupling 
energy $E_J$ over 2000 time steps for 4096 spins with the initial 
conditions of Fig.\ref{smax}. 
(b): Numerical (points) and thermodynamic (line, see section 4.4) 
results of $E_a/E_J$ after long integration times.  
}
\label{eaejt}
\end{figure}

The transition from the almost regular dynamics 
to the irreversible process during the merging of peaks  
is reflected in the share of the total energy of the two parts of the 
Hamiltonian. 
The coupling energy ${\cal H}_J=\sum (1-{\bf S}_n{\bf S}_{n+1})$ 
results from spatial inhomogeneities 
within each of the two domains and from the domain walls 
between them. The anisotropic energy ${\cal H}_a=\frac{1}{2}\sum (1-S_{nz}^2)$ 
depends on the distance of the spins from the poles. 
Fig.\ref{eaejt}a shows the transfer of energy between the two parts of the 
Hamiltonian: 
The initial state contains no coupling energy and the 
anisotropic term is the only contribution. 
Some of this energy flows to the coupling part during the formation 
of the spatially periodic pattern. 
This process is reversed while the system approaches the homogeneous 
state again so that energy is exchanged periodically between ${\cal H}_a$ and 
${\cal H}_J$ (see the periodic behavior for short times in Fig.\ref{eaejt}a). 
\\
The transfer of energy from the anisotropy to the coupling becomes 
irreversible when the humps fuse into compound peaks. The share of coupling 
energy increases even more when compound peaks merge and an increasing 
number of spins overcomes the equator and settles down near to the south pole. 
In phase space, this process is related to Arnold diffusion. The trajectory 
disappears from the initial critical torus and explores more and more 
of the phase space. Finally, the system reaches an equilibrium state 
where the proportion of ${\cal H}_a$ and ${\cal H}_J$ saturates 
(Fig.\ref{eaejt}b). 
The bulk contribution to the coupling energy is due to the inhomogeneity 
within the north-pole domains and not the contribution from the domain walls. 
\subsection{Modified initial conditions}
The systems energy and magnetization per spin given by the initial conditions 
determines the number of spins that point down after a long time. 
We study the final state for initial conditions with varying energies and 
with a modified magnetization  profile. This gives strong numerical 
indications that the two integrals of motion (energy and the magnetization) 
are the key quantities that influence the final state. 
\subsubsection{Varying energies}
\begin{figure}[htb]
\epsfbox{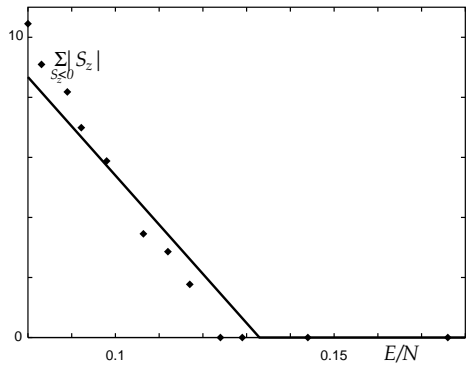}
\caption{  
Numerical (dots) and analytical (line) results for 
the total magnetization $\sum_{n(S_{nz}<0)}|S_{nz}|$ of the spins in the 
southern hemisphere as a function of the total energy $E_a+E_J$ per spin 
for the system of Fig.1a. 
The initial conditions are $S_z=\sqrt{0.84}$, 
$S_{xn}+iS_{yn}=0.4 \exp(ikn)$ with various wavenumbers $k$; 
waves with wavenumbers from $k=0$ to $k=\pi$ provide different amounts 
of coupling energy $E_J$ initially. The spin chain is integrated 
over 100000 time steps. 
For energies per spin below the threshold 
(equation (\ref{mising})), some of the spins condense 
near to the south pole. 
}
\label{trans}
\end{figure}

Spin-wave like initial conditions
$S_{xn}+iS_{yn}=\sqrt{1-S_z^2}\exp(ikn)$ with a given amplitude 
provide energies 
$E=N(1-S_z^2)(\frac{1}{2}+J(1-cos k))$ depending on the 
wave number $k$ while the magnetization $M=NS_z$ is $k$-independent. 
The initial condition $k=0$ in the simulation 
described in Fig.\ref{sprofile}(a),(b) corresponds to the 
minimal energy $E=\frac{N}{2}(1-S_z^2)/2$ that is possible 
for a given magnetization. 
The maximum energy $E=N(1-S_z^2)(\frac{1}{2}+2J)$ corresponds to an 
excitation at the boundary of the Brillouin zone $k=\pi$ 
and any energy between these values can be obtained by a suitable spin wave. \\
We find that some of the spins condense near to the south pole eventually 
only if the energy is below a certain threshold $E_{eq}$ within 
this range. 
Fig.\ref{trans} shows the magnetization of the south-pole 
condensate as a function of the systems energy: 
\begin{enumerate}
\item[{\bf (i)}] 
{\bf Oversaturated phase:} 
Below the energy threshold $E_{eq}$, the number of down-spins is 
proportional to $E_{eq}-E$. The behavior in this is very similar 
to the scenario described before. 
The spatially homogeneous initial condition in the above 
simulations just leads 
to the highest possible proportion of the south-pole condensate. \\
\item[{\bf (ii)}] 
{\bf Overheated phase:} 
For high energies $E>E_{eq}$, no spins are flipped 
down so that 
this magnetization is zero. There is no south-pole condensate 
beyond this threshold; all spins end up in small fluctuations \
near the north pole. 
\end{enumerate}
\subsubsection{Modified magnetization profile}
We have seen that only long-wavelength fluctuations create peaks. 
In contrast, high-energetic initial conditions with short wavelengths 
melt away such peaks of 
spins deviating significantly from the north pole. 
This occurs for an initial condition of a small-amplitude 
$k=\pi$ spin-wave (corresponding to the maximum energy 
in Fig.\ref{trans}) where the spins within a small domain are flipped to the 
southern hemisphere. Fig.\ref{smax}c shows 
the destruction of such a domain in a bath of $k=\pi$ waves. The system 
ends up in an irregular state where all spins are near to the north pole. 
\subsection{Modified equations of motion}
The scenario we have described is widespread in dynamical systems 
and not a specific feature of the Landau-Lifshitz equation. 
A comparison of this scenario with self-focusing 
in related systems indicates that the 
main conditions for the emergence of coherent structures are 
\begin{description}
\item[(i)] the low-amplitude dynamics is governed by an NLS-type of equation, 
\item[(ii)] the system is nonintegrable, 
\item[(iii)] there are two integrals of motion. 
\end{description}
The first of these points basically characterizes the dynamics 
of nonlinear dispersive systems on long scales. The focusing DNLS 
is the system most closely related to the spin chain. 
Also the case of a defocusing nonlinearity will be considered. 
The importance of nonintegrability will be shown by 
the comparison to the integrable Ablowitz-Ladik discretization of the 
NLS equation. On the other side, we will study a system 
with broken rotational symmetry that only conserves the Hamiltonian. 
\subsubsection{Defocusing equation}
The formation of coherent structures in discrete 
nonintegrable systems is not an 
exclusive property of 'focusing' types of discrete NLS or Landau-Lifshitz 
equations. 
The Landau-Lifshitz equation 
$\dot{\bf S}_n={\bf S}_{n}\times ((J({\bf S}_{n-1}+{\bf S}_{n+1})
-{\bf S}_{nz}{\bf e}_z)$ 
with a negative ('easy-plane') anisotropy 
corresponds to the defocusing discrete NLS equation. 
For weak coupling constants 
($J=0.1$), short wavelength ($k=\pi$) 
initial conditions produce coherent structures with spins condensing 
in the equator region where the anisotropic energy has now its minimum. 
Fig.\ref{smax}d shows this 
weak focusing process for the spin chain.
\subsubsection{Discrete NLS equation}
\begin{figure}[htb]
\epsfbox{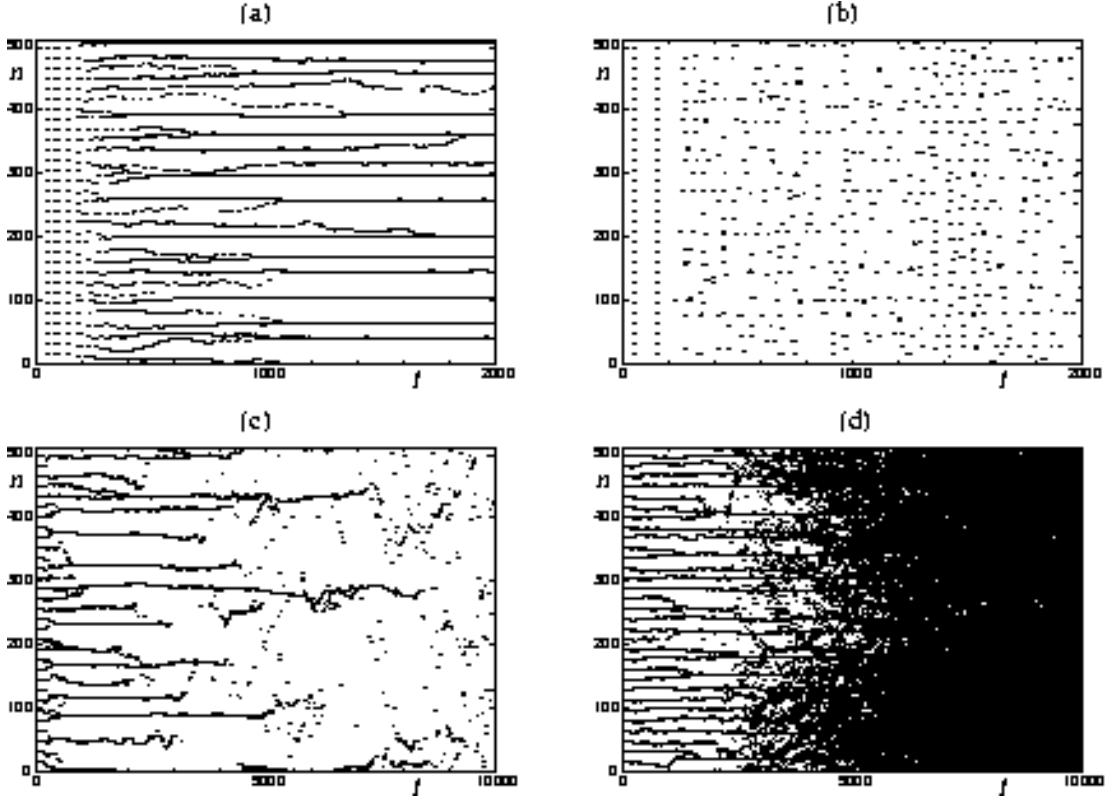}
\caption{  
Integration of the various versions of the discrete 
nonlinear Schr\"odinger equation 
$(J=0.4)$ with 512 lattice sites and periodic boundary conditions. 
The inital conditions are $\phi_n=0.2$ plus noise, 
lattice sites with $|\phi|>0.25$ are marked with dots. 
(a): Integration of the DNLS over 2000 time steps; 
(b): Integration of the of the integrable version of the DNLS over 
2000 time steps. 
(c): DNLS with a contribution $-2\phi_n$ and a symmetry breaking field 
$0.2\phi'_n$ over 10,000 time steps; 
(d): DNLS with a symmetry breaking contribution 
$0.02\phi'_n$ over 10,000 time steps. 
}
\label{nlsmax}
\end{figure}

The focusing nonintegrable DNLS 
$i\dot\phi_n=J(\phi_{n+1}+\phi_{n-1}-2\phi_n)+2|\phi_n|^2\phi_n$ 
has the properties (i)-(iii) just like the Heisenberg spin chain. 
Fig.\ref{nlsmax}a shows the spatiotemporal pattern for the DNLS 
of lattice sites with high amplitudes that is very similar to the 
one described in 
section 3.1 (Fig.\ref{sprofile}a). Again, an unstable periodic pattern 
emerges from a phase-instability of the homogeneous state 
that is unstable itself. \\
The first and the second phase instability are well-known as 
direct consequences of (i) and (ii). 
The second phase instability has been studied in detail 
in the context of NLS equations. 
In phase space, it is related to degenerate tori with less 
than the maximum dimension. Such critical tori exist in integrable 
as well as in nonintegrable systems; they may be stable or unstable. 
The stability of these tori is related to double 
points in the spectral transform \cite{mls}. 
The spectrum of the Lax-operators has been analyzed 
both for an integrable and a nonintegrable version of the discrete 
NLS equation. \\
The fusions of humps lead to peaks with high 
amplitudes (Fig.\ref{nlsmax}a) and finally high-amplitude xenocrysts
emerge from a low-amplitude turbulent background. 
The whole process 
is very similar to the one of 
the Landau-Lifshitz equation (Fig.\ref{smax}a). 
The correspondence persists even in a domain 
where the additional nonlinear terms in the Landau-Lifshitz equation are 
not small. The DNLS-peaks may have different heights while the spin-peaks 
are always south-pole states. \\
The radiation of low-amplitude fluctuations during the merging 
of peaks is related to homoclinic chaos following the break up of 
Kolmogorov-Arnold-Moser tori. 
By discussing the Melnikov-function of the critical tori, \cite{cal}  
have detected homoclinic crossings in the nonintegrable NLS. 
The horseshoe of the homoclinic crossings creates 
the disorder following the fusion of two peaks. \\
A phenomenon similar to the merging of peaks was found 
in a continuous nonintegrable NLS equation \cite{dya}. 
Unlike solitons in integrable systems, collisions of solitary 
solutions of nonintegrable equations 
lead to a transfer of power from the weaker soliton to the stronger 
one while low amplitude waves are radiated. The resulting two-component 
solution contains a decreasing number of growing solitons  
immersed in a sea of weakly turbulent waves. 
In continuous systems, energy is drained by 
infinitesimal scales while the spatial discretization defines a minimal 
length scale. 
\subsubsection{Integrable discrete NLS equation}
Homoclinic chaos as a source of radiation is absent in integrable systems. 
The comparison with the integrable 
discrete NLS 
$i\dot\phi_n=J(\phi_{n+1}+\phi_{n-1}-2\phi_n)
+|\phi_n|^2(\phi_{n-1}+\phi_{n+1})$ 
shows this implication of the nonintegrability (ii). 
The integrable NLS equation exhibits the 
primary phase instability, but not the fusing of neighboring peaks.  
While the dynamics is similar to the nonintegrable system 
initially, the peaks do not 
merge (Fig.\ref{nlsmax}b). Consequently, 
no coherent structures evolve
and the system does not settle down in a disordered equilibrium state.  
The quasiperiodic appearance of humps with relatively low amplitudes 
reflects the phase space structure of nested tori. 

\subsubsection{Particle nonconserving equation of motion}
While the irreversible focusing process is a consequence of 
nonintegrability, it also depends on the 
existence of some remaining integrals of motion. 
The generation of coherent structures 
is very sensitive to perturbations 
that destroy one of the remaining integrals. 
The property (iii) may be changed by breaking the rotational symmetry 
with the contribution $\epsilon Re(\phi_n)$ in the NLS equation.  
The equation is still of Hamiltonian type, 
but the modulus-square norm $\sum|\phi_n|^2$ is not conserved. 
Additional contributions 
$\sim\omega \sum |\phi_n|^2$ to the Hamiltonian and the corresponding term 
$\sim\omega \phi$ in the equation of motion 
$i\dot\phi_n=J(\phi_{n+1}+\phi_{n-1}-2\phi_n)+
\omega \phi_n +\epsilon Re(\phi_n)+
2|\phi_n|^2\phi_n$ 
are now relevant for the dynamics (in the symmetric case, this term just 
describes the same dynamics in different rotating frame systems; 
in the symmetry broken system, the external field $\epsilon$ is 
stationary in the system that rotates with the frequency $\omega$). 
Depending on the sign of $\omega$, two different scenarios are 
observed: 
\begin{description}
\item[{\bf $\omega< 0$:}]
The onset of self-focusing for small times is similar to the symmetric case. 
However, the peaks emerging from the fusing process disintegrate 
eventually into small amplitude fluctuations (Fig.\ref{nlsmax}c). 
\item[{\bf $\omega\ge0$:}]
The onset of the focusing process is again similar to the one 
with particle conservation, but after about 5000 time steps growing 
amplitude fluctuations lead to a disordered state. Unlike the rotationally 
symmetric system, high particle densities are not confined to small 
islands in a sea of low particle density fluctuations. The nonconservation 
of the particle number leads to high (but finite) 
particle density fluctuations everywhere (Fig.\ref{nlsmax}d). 
\end{description}
In the spin chain, similar effects can be reached with an external 
magnetic field that is perpendicular to the anisotropy axis ${\bf e}_z$ 
and an additional z-field. The Hamiltonian now contains the additional 
Zeeman terms $\epsilon S_x + \omega S_z$. 
Due to the broken rotational symmetry the total 
magnetization is no longer an integral of motion. 
\subsection{The leapfrog-discretization of the 
Korteweg-deVries equation}
Iterations of the leapfrog-discretization of the Korteweg-deVries equation 
exhibit a scenario of merging peaks that is very similar to the one found 
in NLS or spin equations. 
However, the leapfrog system undergoes a rapid unbounded growth similar to 
the blow-up in twodimensional NLS systems. 
While the Heisenberg spin chain has a 
well-defined equilibrium, 
the leapfrog system allows us to study the conditions for 
blow-ups in constrained systems. 
We study this phenomenon 
for two initial conditions, the $k=2\pi/3$ mode with a strong correlation 
of $u$ and $v$, and for white noise with no correlation of $u$ and $v$. 
The system consists of 1020 lattice sites with periodic boundary conditions. 
\subsubsection{Correlated initial conditions}
The monochromatic wave with the wavenumber $k=2\pi/3$ 
as initial condition yields 
an exact but phase-unstable \cite{clhe} solution of the 
leapfrog-iteration (\ref{kdvdisc}a) for a sufficiently small 
step size. 
\begin{figure}[htb]
\epsfbox{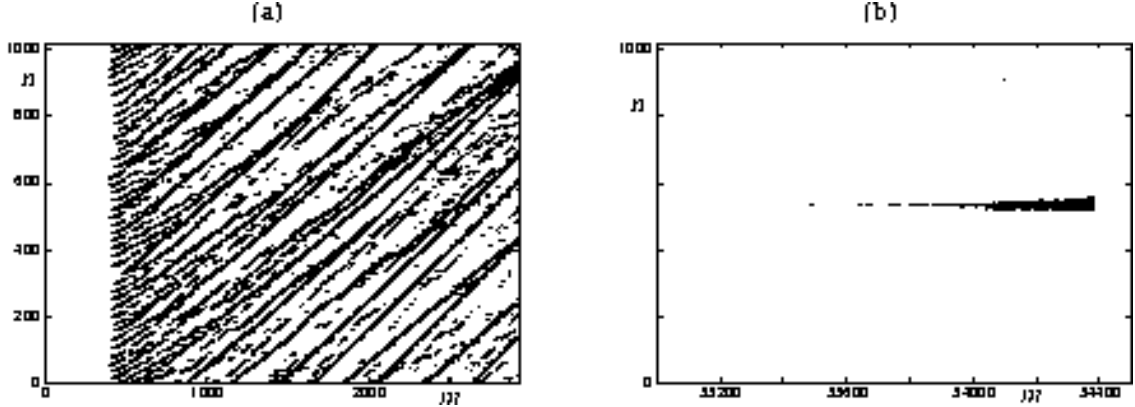}
\caption{  
Iteration of the leapfrog-discretized KdV-equation (\ref{kdvdisc}) 
for a $k=2\pi/3$ wave (a) and for white noise 
initial conditions (b) 
with periodic boundary 
conditions. The amplitude of $u(n)=v(n)$ is $0.1$ initially in (a). 
Locations with high amplitudes (the sum of $u(n)^2$ over five subsequent 
steps is greater than 0.1) are marked with a dot. 
In (b), the threshold 
of $\sqrt{u^2+v^2}$ is 
is $0.045$ while the noise level $\sim 0.01$. 
}
\label{kdvmax}
\end{figure}
This mode is particularly relevant for a stability analysis since it is 
the fastest growing mode for a step size $\tau>2/(3\sqrt{3})$. 
Setting $u_1(n)=v_1(n)$ provides the maximum  
correlation of $u$ and $v$. 
The pattern of peaks Fig.\ref{kdvmax}a  
(lattice sites with high $u(n)^2+v(n)^2$) 
emerges in a manner similar to the spin- and NLS-systems (Fig.\ref{smax} 
and Fig.\ref{nlsmax}): \\
\begin{enumerate}
\item[{\bf (i)}] {\bf Regular behavior: } 
The initial low-amplitude wave is below the threshold 
to be traced in Fig.\ref{kdvmax} for $m<400$. 
\item[{\bf (ii)}] {\bf Merging humps: } 
A phase instability of the initial $k=2\pi/3$ wave leads to 
a modulational pattern with a wavelength of about 20 lattice sites 
that reaches the threshold at $m\approx 400$ so that a spatially periodic pattern 
emerges for $400<m<600$. These periodic humps are wave packets of 
the initial short wave moving towards higher $n$. 
Similar to the spin- and NLS-systems, this pattern itself is slowly modulated. 
The humps approach each other and merge so that a decreasing number 
of peaks of increasing intensity survive. 
Solitary solutions that are high and fast sweep away slower ones. 
The peaks speed and amplitude increase while the width decreases during 
this process. 
They accumulate high amounts of the conserved quantity $<uv>$ just 
like the spin-down domains gather magnetization. 
Fig.\ref{uvprofile} shows the cumulated conserved 
\begin{figure}[htb]
\epsfbox{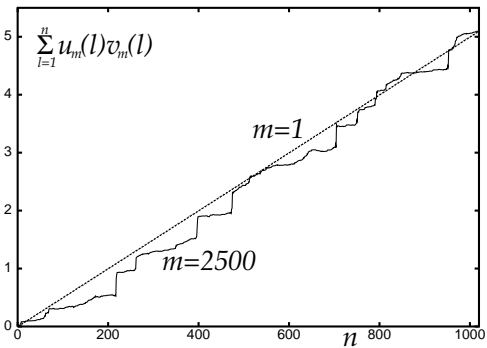}
\caption{  
Cumulated energy  $\sum_{l=1}^n u_m(l)v_m(l)$ 
for the simulation of Fig.\ref{kdvmax} and Fig.\ref{kdvee}
as a function of the lattice site $n$ at the time steps 
$m=1$ and $m=2500$
}
\label{uvprofile}
\end{figure}
quantity $\sum_{l=1}^n u_m(l)v_m(l)$ 
as a function of $n$ (for $n=N$, it is conserved) 
at the beginning ($m=1$) and at 
$m=2500$ when the solitary waves have developed. 
While the conserved correlation $\sum u(n)v(n)$ 
is equally distributed 
in space initially, the formation of solitary 
wave packets gathers an increasing amount of the 
correlation in small xenochrysts immersed in uncorrelated low-amplitude 
fluctuations. 
Up to this point, 
the process is very similar to the self-focusing scenario 
presented in the spin- and NLS-systems. 
\item[{\bf (iii)}] {\bf Rapid divergence: } 
However, despite the fact that for a long time the system appears to 
reach a statistically stationary state, in the end it is clear that no 
equilibrium is attained and the local amplitude rapidly diverges. 
At $m\approx 2960$, two peaks merge at $n\approx 420$ creating an 
all-time high of the amplitude that apparently 
exceeds a certain threshold locally. 
This highest peak now starts to grow rapidly so that the iteration is 
derailed within a few time steps. The features of this rapidly growing 
'monster'-solution will be described in the next section. 
\end{enumerate}
Unlike the modulus square norm of the continuous KdV equation, 
$\sum u(n)^2$ is not conserved in the whole process. 
Fig.\ref{kdvee}a shows $\sum (u(n)+v(n))^2$ and $-\sum (u(n)-v(n))^2$ 
\begin{figure}[htb]
\epsfbox{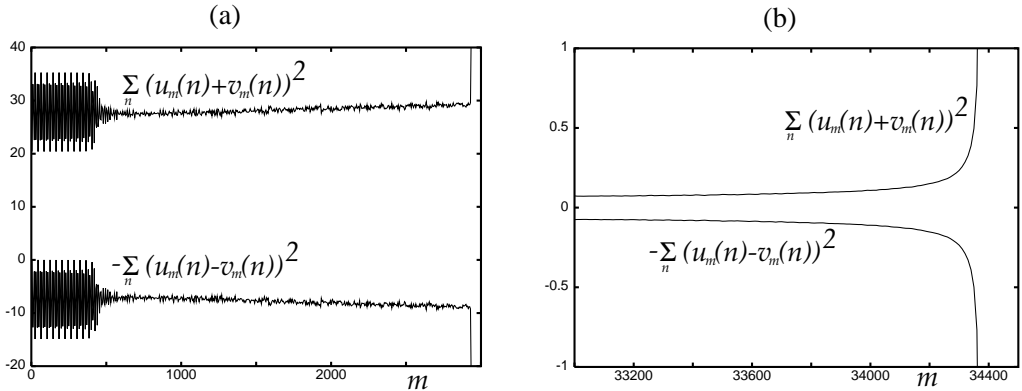}
\caption{  
$\sum(u(n)+v(n))^2$ and $-\sum(u(n)-v(n))^2$
of the leapfrog-discretized KdV system as a function of time $m$ 
for the $k=2\pi/3$-wave (a) and for the white noise initial 
condition (b). 
}
\label{kdvee}
\end{figure}
as a function of the time $m$. $\sum (u(n)+v(n))^2$ equals $4\sum u(n)v(n)$ 
initially while $-\sum (u(n)-v(n))^2$ starts at zero; the sum of both 
quantities $\sim <u(n)v(n)>$ is conserved.  
In the 'reversible' range $m<400$ (i), both quantities 
increase and recur 
to their initial value periodically. 
As the merging of peaks starts (ii), they settle to almost constant values 
undergoing only a very slow increase. This behavior resembles 
Fig.\ref{eaejt} up to the final blow-up (iii) where both quantities 
diverge rapidly. \\
Initial condition with a low amount of $<uv>$ lead to solitary 
solutions that do not reach the threshold for the blow up. 
Their growth ends when a few of them have absorbed 
this conserved quantity and move with the same speed. 
This state however is also unstable because of an  instability 
to be described in the next section. 
\subsubsection{Uncorrelated white noise initial conditions}
Uncorrelated low-amplitude white noise initial conditions $<uv>=0$ lead 
a creeping nonlinear process that suddenly ends up in the same 
sort of local rapid divergence of the amplitude. 
The time elapsing until 
the system blows up is inversely proportional to the square of the 
noise amplitude. For random white noise initial conditions with the 
same amplitude it is Poisson-distributed. \\
Fig.\ref{kdvmax}b shows the spatiotemporal pattern of 
locations where the amplitude slightly exceeds the noise level. 
Unlike the correlated case, 
there is no spatiotemporal pattern of merging peaks. A change in 
the amplitude profile of the noise background 
is hardly detectable even shortly before the blow-up occurs. 
Again three phases of the dynamical behavior can be distinguished:
\begin{enumerate}
\item[{\bf(i)}] {\bf Regular behavior: } 
For about 32000 time steps, the amplitudes are at the level of the noise 
imposed by the initial conditions. During this time, physical structures 
such as solitons can be simulated reliably when they are imposed by the initial
conditions. 
\item[\bf{(ii)}] {\bf  Creeping focusing process}
A weak non-moving maximum emerges at $m \approx 33500$, 
$n\approx 550$ and grows slowly. 
Fig.\ref{kdvmonster} shows the low-pass filtered amplitudes of 
\begin{figure}[htb]
\epsfbox{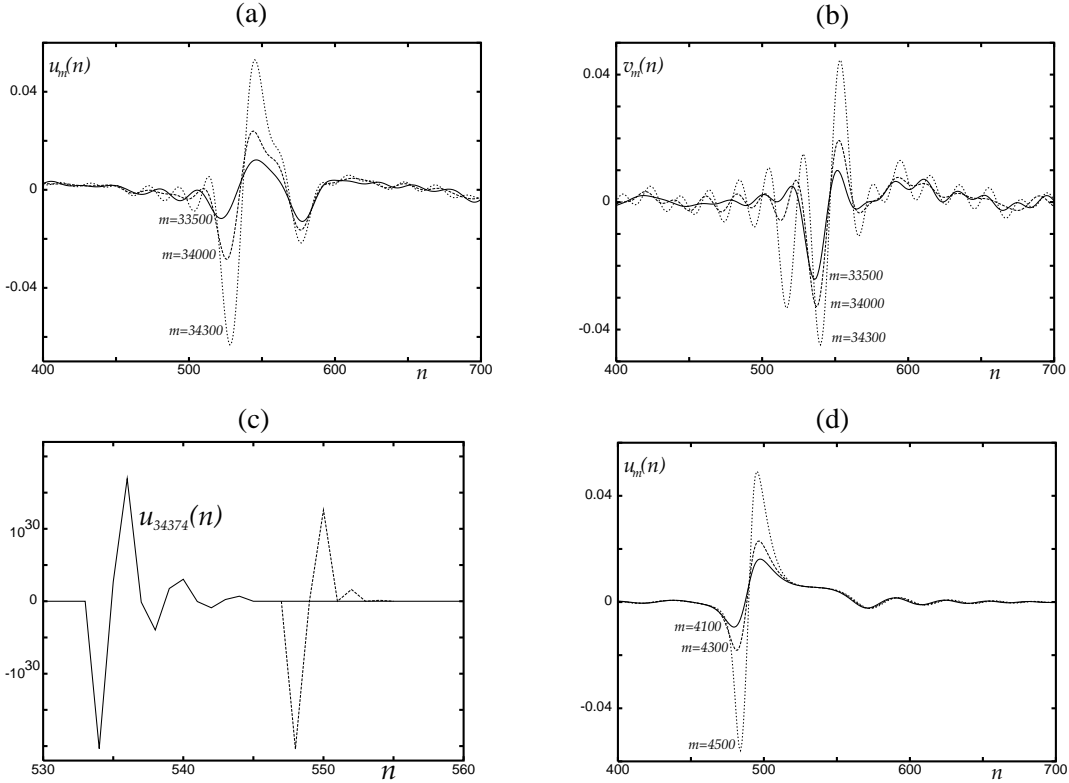}
\caption{  
Profile of the peak evolving out of white noise (Fig.\ref{kdvmax}b). 
Low-pass filtered amplitudes 
$u(n)$ (a) and $v(n)$ (b) at $m=33500$, $m=34000$, 
$m=34300$. 
(c): $u(n)$ at $m=34374$ two time steps before the iteration 
breaks down. The dotted line is the analytical solution. 
(d): $u(n)$ at $m=4100$, $m=4300$ and $m=4500$ for smooth initial 
conditions $v_0(n)=0$, $u_0(n)=0.01/\cosh^2(n-510)$ is very similar to 
the simulation (a). 
}
\label{kdvmonster}
\end{figure}
$u(n)$ (a) and $v(n)$ (b) of this structure at $m=33500$, 
$m=34000$, $m=34300$. 
The low pass filtered amplitude changes significantly with 
each time step, but very little with two subsequent time steps. 
Fig.\ref{ftmonster} shows the corresponding spectral density. 
\item[{\bf(iii)}] {\bf Rapid divergence}
After a slow growing process over about 1000 
time steps this structure suddenly starts  
to grow rapidly and derails the iteration. Fig.\ref{kdvmonster}c 
shows $u(n)$ for this solution at $m=34374$. 
Fig.\ref{ftmonster} shows the spectral density of $<uv>$ at 
the time steps of Fig.\ref{kdvmonster}. 
\begin{figure}[htb]
\epsfbox{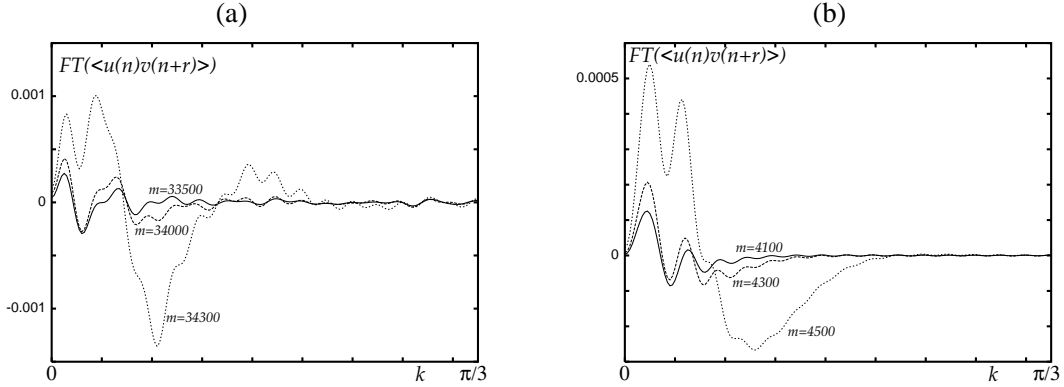}
\caption{  
Spectral density (defined as the Fourier-transform of 
the correlation $<u(n)(v(n+r)+v(n-r))+v(n)(u(n+r)+u(n-r))>$, $r\le 100$ 
for the simulation of Fig.\ref{kdvmax}b at the time steps of 
Fig.\ref{kdvmonster}. (a) shows the slowly growing solution that 
leads to the monster.
(b): The correlation for the smooth initial conditions of 
the simulation \ref{kdvmonster}d.
}
\label{ftmonster}
\end{figure}
This solution (which we call the 'monster' solution because its 
spatial structure resembles the Loch Ness monster with 
several undulations of its tail sticking out of the water) 
appears to be the systems canonical trajectory towards infinite amplitudes. 
\end{enumerate}
Monsters are strongly localized: 
Left to the highest negative amplitude, the head of the monster, 
the amplitudes are close to zero. Right to the head, the amplitude 
of the zig-zag tail decreases rapidly. The monster moves to the left 
by one lattice site with each time step. There are also monsters 
that move to the right; their shape is related to left-moving monsters as 
$u(n)\rightarrow -u(-n)$. 
Most significantly, the amplitude of the monster is squared 
by every time step so that the monster grows as $exp(exp(m))$. 
Fig.\ref{kdvmax}b  
shows a delta-shaped broadening zone of high amplitudes near the blow-up 
since the tail is growing while the head moves towards small $n$. 
\\
This solution can be calculated analytically. 
For a state $u_m(n)=A a(n)$ with a high amplitude $A$, the 
linear terms of equation (\ref{kdvdisc}) may be neglected. 
We assume that the structure grows quadratically and moves to the right 
as $u_{m+1}(n)=2\tau A^2 a(n-1)$. 
Setting $a(n)=0$ for $n>0$ and for negative odd $n$, 
we get the solutions $a(0)=1$, 
$a(-2)=(1-\sqrt{5})/2$, 
$a(-2n-2)=(1-\sqrt{1+4a(-2n)^2})/2$ 
as the solution of $a(n-1)=a(n-1)^2-a(n+1)^2$. 
The right moving solution is obtained by setting $a(n)\rightarrow -a(-n)$. 
The dotted line in Fig.\ref{kdvmonster}c shows this solution where the 
monster's head is submerged. 
\\
Most significantly, the conserved quantities $<u>$, $<v>$ and 
$<uv>$ are zero for this solution; the monster needs no external 
food source for its growth. On the other side, it rapidly produces 
high amounts of $<u^2+v^2>$ (Fig.\ref{kdvee}b).\\
A blow-up with 
these characteristic features follows from 
various initial conditions. 
In the previous section we have described its evolution 
out of solitary solutions that emerge from a Benjamin-Feir type 
of instability and exceed 
a certain threshold after the merging process. 
It can also grow out 
of the noise between these solitary solutions or out of weak white noise 
through a much more dramatic type of instability.

\section{Statistical analysis of the final state}
In this section we will give a detailed interpretation of these results. 
The merging of peaks corresponds to an Arnold-diffusion process in phase space 
that transfers the trajectory from the initial critical torus to less 
distinctive parts of the shell of constant energy 
$<{\cal H}>=E$ and magnetization $<{\cal M}>=M$. 
The numerical findings indicate that the characteristics of the 
final solution are determined by the values of the integrals of motion; 
i.e. that the system reaches a thermodynamic equilibrium.  
One can therefore establish thermodynamic connections between macroscopic 
observables of the final solution and the integrals of motion. 
We will compute the equilibrium statistics of the system and compare 
the results to our numerical findings. 
\subsection{The partition function}
\subsubsection{Low-temperature approximation}
The numerical findings suggest that for low energies the spins point to 
small regions near to the poles (Fig.\ref{spinsketch}a) 
and avoid the region nearer to the equator. 
With $\sigma_n=\pm 1$ one may approximate spins near to the 
north- or south pole as  
\be{sigdef}
S_{nz}\approx\sigma_n(1-\frac{1}{2}(S_{nx}^2+S_{ny}^2))
\ee
Low amplitude fluctuations near the north pole are represented 
by $\sigma_n=1$ and small values of $S_{nx/y}$. The coherent structures 
with spins near to the south pole correspond to $\sigma_n=-1$. 
This matching height of all peaks is 
the main technical advantage of the spin chain. 
Assuming that $\sigma_n\sigma_{n+1}=1$ holds for almost all $n$
(i.e. the number of domain walls is small), 
one may approximate 
$\sum\sigma_n\sigma_{n+1}S_{nx/y}^2\approx \sum_nS_{nx/y}^2$. 
If all spins are close to the poles, the approximate Hamiltonian 
\be{hamiswav}
\begin{array}{cccl}
{\cal H}_{eff}&=&\sum_n&\frac{1}{2}(S_{nx}^2+S_{ny}^2)+J(S_{nx}^2+S_{ny}^2)\\
&&&-J(S_{nx}S_{n+1x}+S_{ny}S_{n+1y})\\
&&&-J\sigma_n\sigma_{n+1}
\end{array}
\ee
represents  a chain of coupled harmonic oscillators 
$(S_{nx},S_{ny})$ 
and a chain of Ising spins $\sigma_n$. 
The approximation holds if the energy is low and the magnetization 
is close to the maximum, i.e. $\sum(1-S_{nz})/N\ll 1$. It 
neglects the coupling between the oscillators $S_{nx/y}$ 
and the Ising spin $\sigma_n$ so that the Hamiltonian splits up into 
a spin-wave Hamiltonian ${\cal H}_w(S_{nx},S_{ny})$ and an Ising Hamiltonian 
${\cal H}_I(\sigma_n)$. ${\cal H}_w$ contains the lowest order terms of $S_{nx/y}$ 
and neglects anharmonic energy contributions of 
neighboring spins that point to the same hemisphere. 
${\cal H}_I$ accounts for the coupling between up and down spins. 
In terms of the stereographic projection, ${\cal H}_w$ contains 
the terms that 
prevail for $|\phi|\ll 1$ while ${\cal H}_I$ allows for the contributions 
for $|\phi|\gg 1$. 
The nonlinearity is only reflected in the Ising magnet. 
\\
The magnetization 
as a second integral of motion may be approximated as 
\be{magapprox}
{\cal M}_{eff}={\cal M}_I(\sigma)+{\cal M}_w(S_x,S_y)
=\sum_n\sigma_n-\sum_n\frac{1}{2}(S_{nx}^2+S_{ny}^2)
\ee
if most of the spins point to the north pole. Again this approximation 
neglects 
higher order terms in $S_{nx/y}$ and contributions $\sigma_nS_{nx/y}^2$ 
with $\sigma_n=-1$. 
\subsubsection{Grandcanonical partition function}
The phase space surface on which 
${\cal M}$ and ${\cal H}$ are constant can be computed most easily 
using the grand partition function 
\be{grancan}
y(\beta,\gamma)=\int e^{-\beta({\cal H}_{eff}-\gamma {\cal M}_{eff})} d\Gamma
\ee
with two parameters $\beta$ and $\gamma$. Unlike the canonical
ensemble \cite{krum}, the grand canonical ensemble  
reflects the second integral of motion by the parameter $\gamma$ 
that controlls the system's magnetization. 
$\beta$ is the inverse temperature while $\gamma$ is an equivalent of a 
magnetic field or chemical potential. 
The exponent in (\ref{grancan}) is a sum 
${\cal H}_{eff}-\gamma {\cal M}_{eff}
=({\cal H}_w-\gamma {\cal M}_w)+({\cal H}_I-\gamma {\cal M}_I)$ 
of a spin-wave contribution 
\be{hspinw}
{\cal H}_w-\gamma {\cal M}_w=
\sum_n \frac{J}{2}((S_{nx}-S_{n+1 x})^2+
(S_{n+1 y}-S_{n+1 y})^2)
+\frac{1+\gamma}{2}(S_{nx}^2+S_{ny}^2)
\ee
that depends only on $S_{nx},S_{ny}$
and an Ising contribution
\be{hising}
{\cal H}_I-\gamma {\cal M}_I=
J\sum_n(1-\sigma_n\sigma_{n+1})-\gamma\sum\sigma_n
\ee
that depends only on $\sigma_n$. 
The partition function (\ref{grancan}) of the whole system is the product 
$y=y_wy_I$ where $y_w$ is obtained by an integration over the variables 
$S_{nx}, S_{ny}$ while $y_I$ is a sum of the configurations of the 
Ising spins $\sigma_n$. 
The grand partition function of the $N$ Ising spins  
\be{yising}
y_I=(cosh(\gamma\beta)+\mu)^N
\ee
with the abbreviation $\mu=\sqrt{sinh^2(\gamma\beta)+e^{-4J\beta}}$ 
is just the canonical partition function of an 
Ising magnet in an external field $\gamma$. 
The linear dynamics of $S_{nx}$ and $S_{ny}$ has the symplectic structure  
$\dot{S}_{nx/y}=\pm\partial {\cal H}_w/\partial S_{ny/x}$ so that a 
phase space volume element 
may be approximated as $d\Gamma=\prod dS_{xn}dS_{yn}$. 
For $\beta\gg 1$, the 
grand partition function of the spin waves 
can be obtained by integrating of $d\Gamma=\prod dS_{xn}dS_{yn}$ 
from minus to plus infinity. Using the abbreviation 
$A=\sqrt{\frac{J}{2}+\frac{1+\gamma}{8}}+\sqrt{\frac{1+\gamma}{8}}$ 
the Gaussian integrals yield 
\be{yspinw}
y_w(\beta,\gamma)=(\frac{\pi}{A^2\beta})^N
\ee
\subsection{Thermodynamic relations}
\subsubsection{Energy and magnetization}
\begin{figure}[htb]
\epsfbox{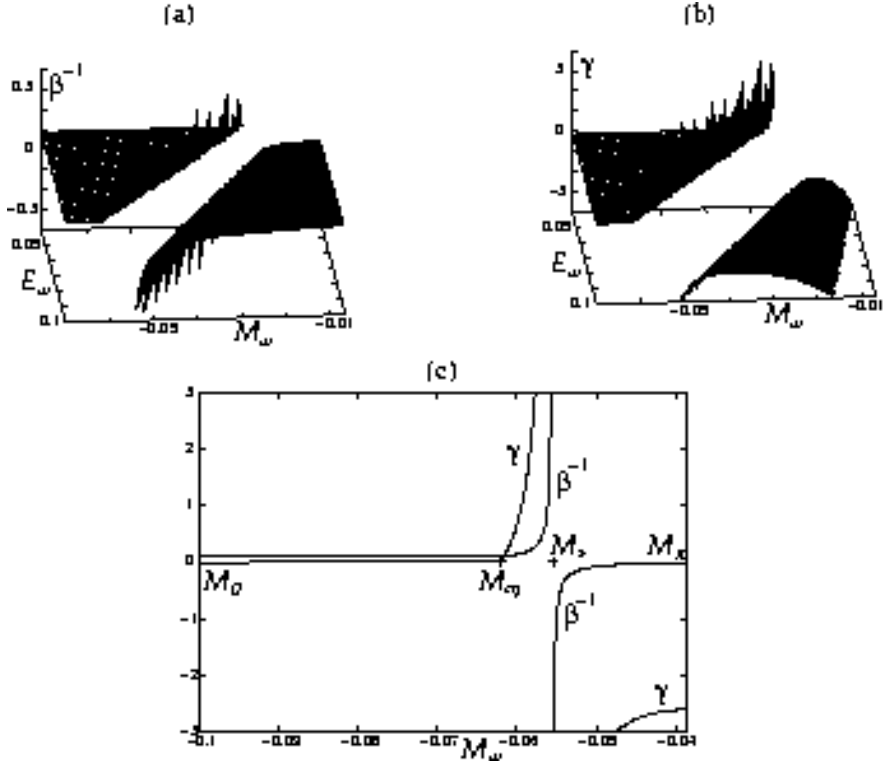}
\caption{  
The temperature $\beta^{-1}(M_w,E_w)$ (a) and the chemical potential 
$\gamma(M_w,E_w)$ (b) as  obtained from the equations (\ref{magpart}) 
and (\ref{hpart}). The energy is fixed as $E_w=0.1$ in (c).
}
\label{thermo}
\end{figure}

The thermodynamic properties of the equilibrium state 
may be derived from the grand partition function  
$ln y(\beta,\gamma)=ln y_w+ln y_I$. 
The parameters $\beta,\gamma$ and the 
conserved quantities  $<{\cal H}>=E$, $<{\cal M}>=M$ are connected by  
\be{magpart}
\begin{array}{ccl}
M&=&M_w+M_I\\
&=&\frac{1}{\beta}\frac{\partial}{\partial \gamma}(ln (y_w)+ln(y_I))\\
&=& N(-\frac{1}{\beta \lambda}
+\frac{sinh(\gamma\beta)}{\mu})
\end{array}
\ee
\be{hpart}
\begin{array}{ccl}
E&=&E_w+E_I\\
&=&(\frac{\gamma}{\beta}\frac{\partial}{\partial \gamma}
-\frac{\partial}{\partial \beta})(ln(y_w)+ln(y_I))\\
&=&\frac{N}{\beta}(1-\frac{\gamma}{\lambda})+
\frac{2NJe^{-4J\beta}}{cosh(\gamma\beta)\mu+\mu^2}
\end{array}
\ee
with $\lambda=\sqrt{4J(1+\gamma)+(1+\gamma)^2}$. 
The partition function is valid for small energies per 
lattice site $E/N\ll 1$ and for small mean deviations 
$1-M/N\ll 1$ of the spins from the north pole. 
Low amplitude initial 
conditions without huge deviations from the 
north pole correspond to magnetizations in the interval 
$1-E/N\le M/N\le 1-E/(N(1+4J))$. The lower bound 
corresponds to a spatially homogeneous initial condition 
while a wave with $k=\pi$ defines the upper bound. \\
$M_I$, $E_I$, $M_w$, $E_w$  are the physically most interesting 
quantities: 
\begin{enumerate}
\item [{$M_I$}]
is the magnetization of the Ising system and measures 
the total extent of the coherent structures. At its maximum $M_I=N$, 
all spins point up while smaller values $N>M_I>M$ indicate the 
existence of coherent structures where the spins point down 
\item [{$E_I$}] is the positive coupling energy of the domain 
boundaries and 
determines the number of coherent structures 
\item [{$M_w$}] is the negative magnetization of small fluctuations 
\item [{$E_w$}] is the positive energy of small fluctuations and comprises a 
coupling term and an anisotropic term 
\end{enumerate}
These quantities can be found by computing   
$\beta$ and $\gamma$ as functions of $E$ and $M$ and then plugging 
$\beta$ and $\gamma$ in the expressions for $E_I$, $M_I$, $E_w$, $M_w$. 
(\ref{magpart}), (\ref{hpart})  
can be solved analytically for the low energy case $E/N\ll 1$. 
The solution (Fig.\ref{thermo}) 
is qualitatively different in the 'oversaturated phase' 
with $M<M_{eq}$ and in the 'overheated phase' $M_{eq}<M<M_{\pi}$ 
with $M_{eq}=N-E/\sqrt{1+4J}$:\\
\subsubsection{Oversaturated phase $M<M_{eq}$:} 
\begin{enumerate}
\item [{\bf (i)}] The temperature $\beta^{-1}\approx E/N\ll 1$ is almost 
independent of $M$\\ 
\item [{\bf (ii)}] The chemical potential 
$\gamma\approx E e^{-2JN/E}/\sqrt{2N(M_{eq}-M)}
\approx 0$ 
is exponentially small unless the magnetization comes very close 
to the transition point where this approximation breaks down. \\
\item [{\bf (iii)}] 
Almost the total energy $E_w\approx E$ is absorbed by the fluctuations 
while the surface energy  
$E_I\sim e^{-2JN/E}\approx 0$ is exponentially small. \\
\item [{\bf (iv)}] 
The fluctuations share 
$M_w\approx -E/(\sqrt{1+4J}N)$ of the magnetization is independent 
of the total magnetization. The remainder of $M-N$ 
is absorbed by the spins that are flipped down. 
The share of spins that is flipped down is at most 
of the order of the energy. 
\end{enumerate}
\subsubsection{Overheated phase $M_{eq}<M<M_{\pi}$: }
\begin{enumerate}
\item [{\bf (i)}] The temperature 
\be{betaofm}
\beta^{-1}=\frac{(E+M-N)(4J(M-N)+E+(M-N))}{(4J(M-N)+2(E+(M-N)))N}
\ee 
\item [{\bf (ii)}] and the chemical potential 
\be{gamma}
\gamma=\frac{(E+M-N)^2}{4J(M-N)^2+2(E+M-N)(M-N)}-1
\ee
both have a singularity at $M_s=N-E/(2J+1)$. 
The temperature is positive between $M_w=M_0$ and the singularity 
because the number of accessible states grows with the energy 
in this range. 
Beyond the singularity, more energy leads to a decreasing number 
of states so that 
the temperature is negative. \\
\item [{\bf (iii)}] The spin waves $E_w\approx E$ again absorb the 
bulk of the energy while  
$E_I\sim e^{-2JN/E}\approx 0$. \\
\item [{\bf (iv)}] 
The Ising-magnetization is near to its maximum 
$M_I\approx N$ independently of $M$. 
Fluctuations contribute the magnetization 
$M_w\approx -E/\lambda$ 
\end{enumerate}
\subsubsection{Transition at $M_{eq}$}
The solution for $M_I$ explains the 
transition behavior of Fig.\ref{trans} (section 3.2.1) 
and the emergence of coherent structures quantitatively. While the 
oversaturated phase corresponds to 
long spin-wave initial conditions, the thermodynamic equilibrium state 
is characterized by coherent structures, i.e. spins pointing 
to the south pole: 
Below the threshold $M_{eq}$, the Ising-magnetization $M_I$ deviates 
significantly from its maximum $M_I=N$ and the number of spins that 
point down increases linearly with $M_{eq}-M$. 
Above the transition the Ising magnetization deviates very little 
from its maximum $M_I=N$, so there are no coherent structures. \\
In both phases almost all energy is absorbed by 
low amplitude fluctuations. 
The surface energy $E_I$ is exponentially small, so that the spins 
form a very small number of domains. The higher number of domains obtained 
numerically indicates that the system does not thermalize completely 
on reasonable time scales. 
\\
As the spins interact only pairwise with a short range in one dimension, 
the transition between the two phases is of diffuse type and not 
a genuine phase transition.   
$\gamma$ and $M_I$ are analytic functions. The slope of $\gamma$ 
increases rapidly within a small interval $\sim e^{-4JN/E}$ 
at $M_{eq}$ so that 
the transition approaches a phase transition as the energy goes to zero. \\

\subsection{The entropy}
\subsubsection{The shape of the entropy function}
\begin{figure}[htb]
\epsfbox{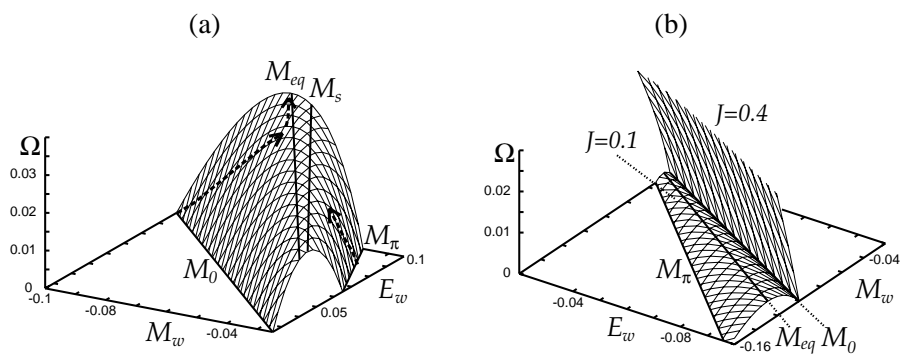}
\caption{  
$\Omega$ for the spin chain as a function of $E_w$ and $M_w$: 
(a) focusing equation with $J=0.4$. The focussing process 
corresponds to the arrow on the left slope, the arrow on the 
crest sketches the merging of peaks. The arrow on the right 
slope represents the destruction of coherent structures by 
short-wavelength fluctuations.  
(b) defocusing case  for $J=0.1$ and $J=0.4$. 
The line $M_{\pi}$ crosses $E_w=0$ for $4J-1=0$.
}
\label{entropyfig}
\end{figure}

The thermodynamic reasons for coherent structures are best 
described in terms of the systems entropy. 
The conserved quantities $M$ and $E$ are known from the initial conditions 
rather than the arguments $\beta, \gamma$ 
of the grand partition function. 
Consequently, the entropy as a function of $M$ and $E$ is the appropriate 
thermodynamic potential. 
The entropy follows from the grand partition function 
by two Legendre transformations  
\be{entropy}
\begin{array}{ccl} 
S&=&ln(y)+\beta(E-\gamma M)\\
&=&(1-\beta\frac{\partial}{\partial\beta})ln(y)
\end{array}
\ee
Both the spin-waves and 
the Ising system contribute to the entropy. The two systems can 
get different shares $E_I$, $M_I$ and $E_w$, $M_w$ 
of the two conserved quantities $E$ and $M$. 
The entropy of the Ising has the form 
$\sim -E_I ln(E_I/N)$. 
The spin-wave entropy per lattice site is given by 
$S_w/N= ln\Omega$
where the total number of accessible microstates is $\Omega^N$ with  
\be{omega}
\Omega=\frac{(E_w+(1+4J)M_w)(E_w+M_w)}{N M_w}
\ee
So the entropy of small fluctuations depends on the energy as 
$\sim N ln(E_w/N)$.
Both systems are coupled thermally, so they have matching temperatures 
$\beta^{-1}$. 
Using $\beta=\frac{\partial S}{\partial E}$ we reestablish the fact 
that the Ising energy $E_I/N\sim e^{-2J\beta}$ is exponentially 
small compared to the 
energy of the fluctuations $E_w/N\sim \beta^{-1}$ for low energies. 
The resulting Ising entropy is again exponentially small 
$S_I\sim e^{-2J\beta}$ compared to the 
spin-waves contribution $S_w\sim -ln\beta+const$. \\
Consequently, the main part of the entropy arises from the 
degrees of freedom of waves with small amplitudes while the 
Ising system only provides an almost constant 
contribution. 
The Ising system can absorb some of the systems  
magnetization without changing its energy and entropy significantly. 
By doing that, the fluctuations share of magnetization can 
also change allowing the fluctuations to maximize their entropy. 
The maximum of $S_w=N ln\Omega$ as a function of $E_w$ and $M_w$ 
is approximately the total entropy maximum. \\
Fig.\ref{entropyfig}a shows the number of states per lattice site $\Omega$ 
of the fluctuations as a function of 
$M_w$ and $E_w$. It has the shape of a crest 
ascending towards higher energies. 
All possible small-amplitude states corresponding  
to positive $\Omega$ are in a triangular region limited by two 
highly ordered solutions ($M_0=-E_w$ and $M_{\pi}=-E_w/(4J+1)$) 
and by the systems total energy $E_w=E$. \\
The area between the lines $M_0$ and $M_{eq}$ again represents 
the oversaturated phase. 
The rim $M_0=-E_w$ with $\Omega=0$ corresponds to a monochromatic wave with 
$k=0$. 
Long wavelength solutions (which are also representative for continuous 
systems) are located at the slope near to this line. 
Such fluctuations have a low ratio $E/(N-M)$. 
Solutions that include high peaks may have even smaller values $M<N+M_0$. 
However, their inert down-magnetized domains have little influence 
on the thermodynamics and these states are similar to the ones at $M_0$. \\
The overheated phase is located between the lines $M_{eq}$ and $M_{\pi}$. 
The rim $M_{\pi}=-E_w/(4J+1)$ with $\Omega=0$ represents a wave $k=\pi$ 
at the boundary of the Brillouin zone. This wave has the 
highest ratio $E/(N-M)$ of all solutions. 
\\  
The systems total energy $E_w=E$ gives a third boundary of the 
accessible states.  The absolute maximum of the entropy is
located on this boundary at $M_w=-E/\sqrt{4J+1}$. 
\subsubsection{Coherent structures}
The formation of peaks (i.e. down-magnetized 
domains) can be understood as the maximization of the entropy under 
the restriction of the conserved quantities. 
Formation and merging or destruction of coherent structures 
is represented by paths from the slopes to the crest 
and to the entropy maximum typifying the phenomena that 
have been observed numerically. While these are nonequilibrium 
processes since $M_w$ and $E_w$ are changing, equation (\ref{omega}) 
gives the equlibrium entropy for a system thermalizing at particular 
constant values of $M_w$, $E_w$.\\
\begin{enumerate}
\item[{\bf(i)}]{\bf Formation of coherent structures in the 
oversaturated phase: }
For $M<N+M_{eq}$ (or $M_w<M_{eq}$, e.g. for 
spatially homogeneous initial conditions or long waves), 
the system can increase $M_w<0$ and decrease $M_I>0$ by 
flipping spins from the north 
to the south. In the entropy profile 
this means that the system is allowed to move 
from the $M_0$-side in the direction towards $M_{eq}$ 
(along the arrow at the left slope in Fig.\ref{entropyfig}a). 
This leads to an increase of the spin-wave entropy 
$S_w$ for initial condition on the $M_0$-side of the slope. 
This process stops when the crest is reached  
so that the ideal amount of magnetization $M_{eq}$ is allocated to the 
spin waves. For long-wavelength initial conditions, the formation of 
coherent structures allows the exploitation of short-wavelength degrees 
of freedom to increase the entropy. \\
An additional increase of the fluctuations entropy may be reached by 
transferring energy from $E_I$ to $E_w$. This happens when merging 
down-magnetized domains reduce the domain-wall energy 
contributing to $E_I$. In Fig.\ref{entropyfig}a we can identify 
the route along the crest $M_{eq}$ with this process. 
Finally, $E_w$ absorbs almost all energy at the summit
leaving little energy $E_I\ll E_w$ for domain boundaries. 
The resulting magnetization of the Ising magnet is 
\be{mising}
M_I=M+\frac{E}{\sqrt{4J+1}}
\ee 
\item[{\bf(ii)}] {\bf Destruction of coherent structures 
in the overheated phase: } 
For $M>N+M_{eq}$ (or $0>M_w>M_{eq}$, e.g. a 
$k=\pi$-spin wave as initial condition), 
flipping spins down is impossible because this would 
decrease the entropy. 
The opposite  movement starting from the $M_{\pi}$ slope towards 
the crest $M_{eq}$ is only possible if some spins are already flipped 
down so that they may be flipped up now. 
This type of thermalization process occurs for initial conditions 
of spin-down xenochrysts immersed in short-wavelength low amplitude 
fluctuations. This process ends if either 
all spins point up ($M_I=N$) or if 
the ideal amount of magnetization $M_{eq}$ is allocated to the spin waves.  
The crest of the entropy may be 
approached from the $M_{\pi}$-side by melting 
existing coherent structures away (arrow at the right slope in 
Fig.\ref{smax}c). 
\end{enumerate}
In Fig.\ref{trans} compares numerical and analytical 
results for the number of down-spins 
as a function of the total energy, while the total magnetization 
is fixed. For low energies (long wave initial conditions), a relatively 
big number of spins points down so that $M_I$ is smaller. For higher energies 
(smaller wavelengths), the number of down spins decreases and reaches zero 
at the transition point. 
In Fig.\ref{entropyfig}a, these initial conditions correspond 
to energies on a line 
$M=const$ connecting points on the lines $M_0$ and $M_{\pi}$. 
The threshold of Fig.\ref{trans} 
corresponds to the intersection point of the line 
$M=const$ and the crest $M_{eq}(E)$. This threshold is 
obtained for any path that crosses $M_{eq}$. 
Below the threshold, the spin-wave entropy can be maximized 
by flipping spins down according to equation (\ref{mising}). 
Above the threshold only an exponentially small 
share of spins point down. Again, the transition is analytic but 
very sharp.
\\
We conclude that the thermalization of energy under the constraint 
of the second 
integral of motion produces high-amplitude peaks emerging from an 
irregular low-amplitude background. 
The formation of coherent structures allows the system to increase its entropy 
of low-amplitude fluctuations by allocating the right amount of magnetization 
to the spin-waves.  
While the total magnetization $M=M_w+M_I$ is constant, 
the spin-wave part $M_w=-\sum(S_{xn}^2+S_{yn}^2)/2$ 
can turn over magnetization to 
the Ising part $M_I=\sum\sigma_n$ that can attain any value in the interval
$M\le M_I\le N$. 
This enhances the entropy of the spin waves 
while the Ising entropy is negligible. 
The entropy increases with $E_w$ so that almost all energy is 
allocated to the fluctuations. This explains the merging process of 
the peaks where energy is transfered from $E_I$ to $E_w$. 
\subsection{Power spectrum and particle conservation}
\begin{figure}[htb]
\epsfbox{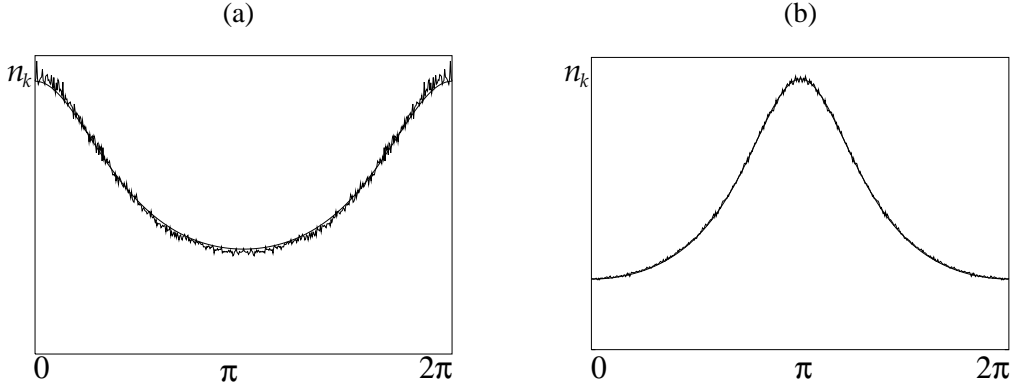}
\caption{  
Spatial power spectra of the spin chain 
averaged over 100000 time steps after a previous 
integration over 100000 time steps and the corresponding   
Rayleigh-Jeans distributions (\ref{power}). 
(a) was obtained for homogeneous initial conditions, 
(b) follows from a noisy short-wave ($k\approx 2\pi/3$) 
initial condition related to negative $\beta$ and $\gamma$.  
}
\label{eqpower}
\end{figure}

Spin waves with a wavenumber $k$ 
contribute the power 
\be{power}
<n_k>=\frac{1}{\beta(2J(1-cosk)+1+\gamma)}
\ee
to the Hamiltonian $E=\sum_k n_k\omega_k$
with $n_k=S_{kx}S_{-kx}+S_{ky}S_{-ky}$ and $\omega_k=2J(1-cosk)+1$. 
The 'particle number' $\sum n_k$ is related to the magnetization 
$M_w$. 
Fig.\ref{eqpower}a compares the Rayleigh-Jeans distribution 
$<n_k>=T/(\gamma+\omega_k)$ 
to numerical simulations. 
\begin{itemize}
\item[{\bf(i)}]
In the overheated phase below the transition, 
the distribution is independent of $M$ 
since $\gamma$ is exponentially small. 
The energy $n_k\omega_k$ is distributed equally over the $k$-space 
while small wavenumbers 
have the highest power $n_k$. 
The power spectrum Fig.\ref{eqpower}a is 
almost unchanged throughout the
oversaturated phase and the fluctuations energy per
particle is constantly $\sqrt{4J+1}$. 
The systems surplus of particles condenses at the south-pole state 
with low energies per particle, i.e. some spins are flipped to the south
pole. The Rayleigh-Jeans distribution is attained during the merging process 
starting from the peak-like spectrum 
of the initial monochromatic wave. 
The fluctuations determine the equilibrium ratio of anisotropic energy 
and coupling energy (Fig.\ref{eaejt}a) as 
$
E_a/E_J=(\sqrt{4J+1}-1)^{-1}
$. 
Fig.\ref{eaejt}b compares this formula to the results of simulations 
with various coupling parameters $J$. Deviations are due to the fact that the 
system does not reach the perfect equilibrium after the integration. 
Some additional domain walls lead to slightly increased values of $E_I$. 
\item[{\bf(ii)}]
Above the threshold, $\gamma$ strongly depends on $M$ and $E$ 
and the power spectrum is deformed. 
The temperature  becomes negative 
in the strongly 'overheated' domain since the entropy as a function of $E$ 
decreases for $M>M_{\infty}$. 
In this range, most of the energy is due to short waves 
(see Fig.\ref{eqpower}b) 
with $\beta^{-1}\approx-0.2$, $\gamma\approx -2.7$). 
\end{itemize}
\subsection{Defocusing equation}
Thermodynamics of the 'defocusing' case (Fig.\ref{smax}d) with a 
negative anisotropy 
${\cal H}=\sum_n J(1-{\bf S}_n {\bf S}_{n+1})-(1-S_{zn}^2)/2$ 
is slightly more complex. 
We obtain $\Omega=(E+(4J-1)M_w)(E-M_w)/M_w$ where $E$ is negative. 
We distinguish two cases: 
\begin{itemize}
\item[{\bf(i)}]
For weak coupling $4J-1<0$, 
$\Omega$ has got a maximum at 
$M_{eq}=E/\sqrt{-4J+1}$ (see $\Omega$-shell for 
$J=0.1$ in Fig.\ref{entropyfig}b). 
$\Omega$ is zero for $M_0\equiv E<0$ and for 
$M_{\pi}\equiv -E/(4J-1)<0$. Differently from the system 
with a positive sign of the nonlinearity in (\ref{spinchain}), now 
$M_{\pi}<M_0$, i.e. short wavelength spin-waves have the lowest magnetization 
for a given energy. Starting from $M_{\pi}$, the system can approach 
$M_{eq}$ and increase its entropy by flipping down single spins. 
In contrast to the process described earlier, it is now favorable 
to store a maximum of energy in the domain boundaries. 
\item[{\bf(ii)}]
For strong coupling ($4J-1>0$), 
$\Omega$ as a function of $M$ decreases in the whole interval of accessible 
values of $M_w<M_0$ or $M_w<M_{\pi}$ 
and is zero at the homogeneous state 
$M_w=M_0\equiv E$ and for short waves $M_w=M_{\pi}$ 
(see $J=0.4$ in Fig.\ref{entropyfig}b). 
The entropy of the spin-waves can not be 
increased by decreasing the magnetization. 
Thermodynamics allows a weak focusing process by storing energy in 
domain walls starting from $M_{\pi}$, but we have found this process 
numerically only in the DNLS system.
\end{itemize}
\subsection{NLS systems versus spin systems}
\begin{figure}[htb]
\epsfbox{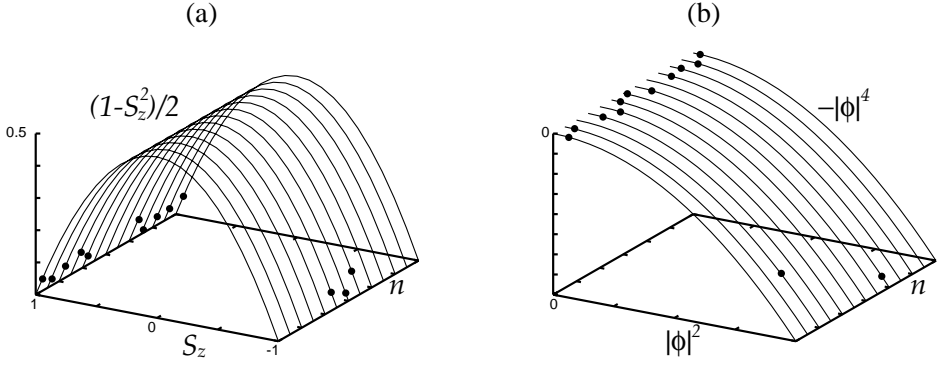}
\caption{  
Sketch of (a) the anisotropic energy $(1-S_{nz}^2)/2$ 
of the spins as a function 
of $1-S_{nz}$ and (b) of the nonlinear energy energy $-|\phi_n|^4$ 
as a function of the particle number $|\phi_n|^2$. The dots indicate 
the typical equilibrium position: Most lattice sites have low 
particle densities while small domains attain high particle densities. 
In the spin chain, these domains extend over several lattice sites. 
}
\label{potential}
\end{figure}

Thermodynamics supports the equivalence of spins and NLS-systems 
with respect to the formation of coherent structures. 
Fig.\ref{potential} shows the nonlinear NLS-energy $-|\phi_n|^4$ 
at lattice sites $n$ as a function of the particle number $|\phi_n|^2$ 
and the corresponding anisotropic spin-energy $(1-S_{nz}^2)/2$ as a function 
of the conserved negative magnetization $1-S_{nz}$; 
the points indicate typical states of the oscillators in equilibrium. 
The spin system has two states with a low 
energy per lattice site: The north pole state is characterized by 
high energies per particle $E/(N-M)$ while the specific energy is low 
near the south pole. The latter state is more fuzzy 
for NLS systems since the potential energy $-|\phi_n|^4$ 
and the particle number $|\phi_n|^2$ per lattice site are unbounded. 
Any contribution proportional to the second integrals may be 
added to the energies; consequently, the ratio of 
the potential energy and the particle number per lattice site is relevant. 
In the NLS-system, 
$-|\phi_n|^4/|\phi_n|^2$ equals 
the negative particle number at the lattice site. 
Similarly, the corresponding ratio $(1-S_{nz}^2)/(1-S_{nz})=2-(1-S_{nz})$ 
decreases linearly with the 'particle number' $1-S_{nz}$. 
The only relevant difference is again the maximum 'particle number' 
$1-S_z$ per lattice site corresponding to the down-spins of 
coherent structures 
while the peak-amplitude of the NLS-system can obtain various values 
depending on the initial conditions. As a result of that, each 
domain with a high amplitude consist of only one lattice site 
that absorbs the particles.\\
Both systems condense some particles in the low-energetic coherent 
structures in order to increase the energy of the remaining particles. 
In both systems the the energy per particle of the condensates 
decreases linearly with the particle density per lattice site. 
This energy becomes available to the coupling part of the Hamiltonian so that 
the system can explore degrees of freedom of wave-like fluctuations. 
This disordered state can only absorb a finite amount of energy per 
lattice site since the lattice constant limits the shortest wave lengths. 
In contrast, fluctuations on infinitesimal length scales of continuous 
systems absorb all the energy  
in the coupling term $\sim|\nabla\phi|^2$ 
as a zero-energy peak absorbs all particles 
so that the system blows up in finite time. 
\subsection{Particle-nonconserving systems}
The separation into a high- and a low-amplitude state 
follows from the entropy maximization 
under the restriction of the second conserved quantity. 
This separation of the system into small fluctuations 
and high peaks does not occur if the second quantity 
is not conserved (section 3.3.4). 
The system can produce or annihilate particles 
to increase the fluctuations entropy 
and thermalizes on the energy shell without further constraints. 
The Hamiltonian 
${\cal H}=\sum_n J(2\phi_n\phi_n^*-\phi_n\phi_{n+1}^*-\phi_n^*\phi_{n+1})
-\omega|\phi_n|^2-|\phi_n|^4$ 
(we neglect the small symmetry breaking term)  
may be considered as a sum $E=T+V$ of a coupling term and a potential term. 
The sign of $\omega$ in $-\omega|\phi_n|^2-|\phi_n|^4$ decides if there 
is a potential well or a maximum at $\phi=0$. 
This explains the two types of thermalization found in section 3.3.4: 
\begin{description}
\item[$\omega< 0$] The trajectories change very little under 
the influence of the weak symmetry-breaking field for small times. 
A pattern of high peaks emerges initially, 
but disappears again since the surplus particles are annihilated 
(Fig.\ref{nlsmax}c). 
The system finally settles into a state of small fluctuations 
trapped in the potential well. However, some oscillators may 
escape from the local energy minimum and attain 
high amplitudes for weaker symmetry breaking fields.
\item[$\omega\ge0$]
The system thermalizes in a state of high-amplitude fluctuations. 
In the discrete NLS-system, the amplitudes continue to grow 
without bound since particles are created Fig.\ref{nlsmax}d. 
Energy is transferred 
from the potential to the coupling term so that $T$ and $|V|$ 
both grow. This growth stops stops finally so that the amplitudes 
remain finite. 
\end{description}
It is the shape of the potential that leads to low-amplitude 
fluctuations by particle annihilation (i) or to particle creation 
allowing high-amplitude fluctuations (ii) by thermalization. 
Interestingly, particle nonconservation does not led to fluctuations with infinite 
amplitudes in (ii); the particle production stops finally. 
The reason for this is the mismatch of the orders of 
the potential energy $V\sim-|\phi|^4$ and the coupling term 
that grows only quadratically with the amplitude. 
Beyond certain high amplitudes the coupling energy can not 
absorb any more energy that is released by the potential. 
A further increase of the amplitude would distort the Rayleigh-Jeans 
distribution towards high wave numbers and reduce the systems entropy; 
particle production has to stop therefore. 
In other words, the energy shell does not contain 
states with infinite amplitudes. \\
The Landau-Lifshitz equation with $\omega\le0$ leads to small 
fluctuations since the anisotropic energy has quadratic minima 
at the poles. 
The potential energy is maximal at the north pole 
for sufficiently strong values $\omega>0$, so that 
large but finite fluctuations emerge. 
\subsection{Leapfrog-discretized Korteweg-de Vries equation}
The intermediate dynamics of 
the discretized KdV equation resembles the spin systems 
formation of an equilibrium state. The blow-up however is   
an intrinsic nonequilibrium process. 
We study the  phase space volume that is accessible to the system 
during this process. 
\subsubsection{Phase space shell}
The leapfrog-discretized KdV-equation (\ref{kdvdisc}) is a nonlinear, 
area preserving mapping in the 
$N$ variables $u(n),v(n)$. 
The area preserving property can be seen 
from the Jacobian 
\begin{equation}\label{jacobi}
J=\left|
\begin{array}{cc}
(\frac{\partial G(u(n),...)}{\partial u(l)})&
I\\
I&
0
\end{array}
\right|
=1
\end{equation}
where $I$ is the identity matrix and $0$ is the zero matrix. 
$G$ comprises the linear and the nonlinear derivative term 
of (\ref{kdvdisc}). 
The phase space that is accessible to the system 
is restricted by the intergals of motion 
$\sum u_m(n)v_m(n)$, 
$\sum u_{2m}(n)=\sum v_{2m+1}(n)$ and 
$\sum u_{2m+1}(n)=\sum v_{2m}(n)$.  
Introducing the variables 
$P_m(n)=u_m(n)+v_m(n)$ and 
$Q_m(n)=u_m(n)-v_m(n)$, the conserved quantity $\sum u(n)v(n)$ 
of (\ref{kdvdisc}) may be written as 
\begin{equation}\label{energi12}
2<uv>=2\sum u(n)v(n)/N=\frac{1}{2N}\sum_n P(n)^2
-\frac{1}{2N}\sum_n Q(n)^2
\end{equation}
as the difference of two unbounded positive terms while the 
nonconserved modulus square norm is 
\begin{equation}\label{modsq}
<u^2+v^2>=\sum (u(n)^2+v(n)^2)/N=\frac{1}{2N}\sum_n P(n)^2
+\frac{1}{2N}\sum_n Q(n)^2
\end{equation}
Solutions of (\ref{kdvdisc}) which correspond to physical 
solutions have $u(n)\approx v(n)$ so that $Q(n)\approx 0$. 
Roughly speaking, $P(n)$ 
is associated to the 
physical modes which are solutions of the original partial 
differential equation 
while $Q(n)$ is associated to spurious computational modes. 
The integrals $<P>=\sum P(n)/N$ and $<Q>=\sum Q(n)/N$ are linked to the 
system's $k=0$ modes. 
It is useful to consider the phase space only at even 
(or equivalently only at odd) time steps since the sign of $<Q>$ 
changes with each time step. 
\\
The modulus-square norm $<u^2+v^2>$ is almost constant on an intermediate 
time scale $500<m<2800$ in Fig.\ref{kdvee}a and until the blow up $(m<34000)$ 
in Fig.\ref{kdvee}b. 
While the phase space shell defined by constant $<u>$, $<v>$ and $<uv>$ 
has an infinite volume, the additional constraint of keeping 
$<u^2+v^2>$ constant gives a finite phase space shell 
$\nu(<u^2+v^2>)$.  The total phase space shell without 
this constraint is given by 
$\Omega\sim\int \nu(<u^2+v^2>) d<u^2+v^2>$. 
\\
The map (\ref{kdvdisc}) is again area preserving in $P(n)$ and $Q(n)$,  
and the microcanonical partition function can be rewritten as 
\begin{equation}\label{microsep2}
\begin{array}{ccl}
\nu(<u^2+v^2>)&\sim&
\int\delta (\sum P(n)^2/2-N<P^2/2>)\delta (\sum P(n)-N<P>)\prod dP(n)\\
&\times&\int\delta (\sum Q(n)^2/2-N<Q^2/2>)\delta (\sum Q(n)-N<Q>)\prod dQ(n) 
\end{array}
\end{equation}
The integrals over $dP(n)$ and over $dQ(n)$ each 
measure the intersection of a hypersphere 
with the radius $<P^2/2>$ (and $<Q^2/2>$) and a hyperplane 
that intersects the $P(n)$-axes 
at $P(n)=<P>$, (and $Q(n)$-axes at $Q(n)=<Q>$), and one finds 
\begin{equation}\label{microsep3}
\begin{array}{ccl}
\nu(<u^2+v^2>)&\sim&(<P^2>-<P>^2)^{N-1}(<Q^2>-<Q>^2)^{N-1}\\
&\sim&((<u^2>+<v^2>-<u>^2-<v>^2)^2-4(<uv>-<u><v>)^2)^{N-1}
\end{array}
\end{equation}
This expression grows rapidly with 
$<u^2+v^2>$. On the other side, it decreases with the correlation 
$<uv>$. 
Since there is no upper limit for the modulus-square norm, 
the surface $\Omega$ of the conserved quantities phase space shell 
is infinite. 
\subsubsection{The focusing process for correlated initial conditions}
During the focusing process Fig.\ref{kdvmax}a the system gathers huge quantities 
of the correlation $<uv>$ and of $<u^2+v^2>$ into dense solitary 
waves so that the remaining space is filled with uncorrelated 
white noise of $u$ and of $v$. It is tempting to interpret 
this phenomenon along the lines of the formation of down-magnetized 
domains that allowed the spin chain to maximize the entropy of 
small fluctuations. The partition function (\ref{microsep2}) 
however gives no reason for this interpretation; 
$<uv>$ may well be distributed 
equally in space in (\ref{microsep3}). The main problem is that 
$<u^2+v^2>$ is not conserved. The system can increase this 
quantity to increase its entropy. This is what happens in 
the first 400 time steps in Fig.\ref{kdvee}a: The initial 
reversible process increases and decreases this 
quantity as the trajectory is close to a homoclinc orbit linked 
to the phase-instability of the initial $k=2\pi/3$-wave. As 
the trajectory disappears from this orbit, $<u^2+v^2>$ reaches 
a plateau above its initial value that increases 
slowly during subsequent mergings of solitons  
until shortly before the blow-up. 
The solitons are highly 
correlated wave-packets of the $k=2\pi/3$ carrier wave, so they 
contain both $<u^2+v^2>$ and $<uv>$.   
The formation of the solitary waves allows the 
system to increase $<u^2+v^2>$ thereby increasing $\nu$. 
\subsubsection{The focusing process for uncorrelated white noise initial conditions}
While nonlinear terms are extremely weak for the initial noise level $\sim 0.01$ (Fig.\ref{kdvmax}), 
there is an insidious nonlinear process 
that increases the amplitude slowly to the level where 
the rapid growth can occur. 
Fig.\ref{kdvmonster} shows 
the low pass filtered amplitude of $u(n)$ (a) and $v(n)$ (b) long before 
the blow up occurs at this site. 
Its shape changes 
only slowly over even (respectively odd) time steps.  
$u_{2m}(n)$, $v_{2m}(n)$ both grow slowly in time, but they 
deviate substantially from each other. Interestingly, 
the feedback from short waves changes this dynamics very little. 
Fig.\ref{kdvmonster}d shows the dynamics for smooth long wave initial 
conditions $u_0(n)=0.01/\cosh^2(n-510)$, $v_0(n)=0$. 
The solution's shape is very similar to the solution emerging 
out of white uncorrelated noise in Fig.\ref{kdvmonster}a and it also 
initiates the 'monster' solution. 
The amplitudes may be described by continuous functions in 
space and time as  
$u_{2m}(n)=r(t,x)$ and $v_{2m}(n)=s(t,x)$.  
Their dynamics is given 
by two coupled continuous KdV equations 
\begin{equation}\label{dkdv}
\begin{array}{ccl}
\dot r&=&s_{xxx}-6ss_x\\
\dot s&=&r_{xxx}-6rr_x
\end{array}
\end{equation}
These equations describe the long-wave dynamics of the leapfrog-system 
where the strength of the physical and the computational mode is of the 
same order. 
Unlike the KdV equation, spatially homogeneous solutions of 
the coupled KdV equations can be phase unstable.  
Constant solutions $r=r_0$ and $s=s_0$ have the eigenvalues 
$\lambda^2=-r_0s_0k^2-(r_0+s_0)k^4-k^6$ where $k$ 
has been rescaled. They are  
unstable for $r_0s_0<0$ and for $r_0=0$, $s_0<0$ 
(or for $r_0<0$, $s_0=0$). The first instability is very similar 
to the Benjamin-Feir modulational instability and leads to 
traveling solitary waves which eventually can grow sufficiently 
large to access the 'monster' solution. 
The second is new and particularly 
interesting as it can initiate the 'monster' solution in a region 
where $<uv>$ is zero. Its unstable mode has a shorter wavelength 
than the Benjamin-Feir mode (although still long compared to the 
lattice constant) and moreover grows faster. 
We argue below that this saddle point in the phase space is accessed readily 
by an evolving solution because of an inverse flux of the power spectrum 
of the $u,v$ correlation towards low wavenumbers. 
Once the system comes close enough to this starting point, 
a localized solution 
grows irreversibly until it reaches the size necessary for the rapidly 
growing monster solution which is a 
heteroclinic connection to infinity \cite{nera}. 
\subsubsection{The blow-up of amplitudes}
As (\ref{kdvdisc}) is nonintegrable, it is hardly surprising 
that the trajectory eventually separates from any orbit 
shadowing a solution of the original partial differential 
equation. The trajectory simply can disappear from the 
original Kolmogorov-Arnold-Moser torus by  
Arnold diffusion and explore regions of the phase space 
that are most likely connected with high amplitudes. 
The amazing finding is that this 
process can lead to such a rapid and unpredictable 
divergence of the amplitude. 
This feature is absent in the spin system where the phase space itself is 
compact. In discrete NLS systems, the 
coupling energy is restricted by the lattice constant and the 
conserved particle number $\sum|\phi|^2$. The first restriction 
is absent in the twodimensional continuous NLS equation, a canonical 
example of a system with finite-time wave collapses. 
The fixed energy is the difference of two energies $\int |\nabla \phi|^2 dV$ 
and $\int |\phi|^4 dV$ that each are not conserved 
and that can attain any value in a half-open interval. 
The corresponding conserved quantity in the leapfrog scheme is 
$2\sum u(n)v(n)=\frac{1}{2}\sum_n P(n)^2
-\frac{1}{2}\sum_n Q(n)^2$ 
where $\sum P(n)^2$ and $\sum Q(n)^2$ each may grow indefinitely. 
This suggests that a blow-up occurs in systems where   
\begin{itemize}
\item[{\bf (i)}] the phase space non-compact 
\item[{\bf (ii)}] the integral of motion constraining the phase space 
is the difference of two positive unbounded quantities
\end{itemize}
We conjecture that the blow-up is the generic way of thermalization in 
such systems. 
The exploration of the phase space shell defined by a constant  
difference of two positive definite energies 
leads to finite-time singularities since both energies 
can grow unrestrictedly at the same rate. 
This may happen if a solitary structure grows beyond a certain 
threshold, or more surprisingly in a sudden eruption out of 
low amplitude fluctuations. 
The analytically known monster solution 
serves as the canonical highway to infinity during the blow-up. \\
Analogous to the collapse in the two dimensional 
nonlinear Schr\"odinger equation this might be an inevitable consequence 
of a condensation process. 
In that context, the spectral energy 
whose density is approximately $\omega_k n_k$ (where $\omega_k=k^2$ 
is the frequency or energy of a wave vector $\vec{k}$ and $n_k$ 
is the particle density or waveaction) has a net flux to high 
wavenumbers. Because both $\sum\omega_k n_k$ and $n_k$ are approximately 
conserved, this means that there must be a corresponding flux of 
particles towards low wavenumbers. This flux leads to the growth of a 
condensate as particles are absorbed at $k=0$. 
For the focusing NLS equation, this condensate is unstable 
and leads to the formation of collapsing filaments. But the instability 
is very robust and in fact the collapses begin 
before $n_k$ is all concentrated at $k=0$. As soon as 
there is sufficient $n_k$ near $k=0$, the collapses begin. \\
Collapses are inevitable because of the finite flux of particle density  
to long waves. In the present context, a similar scenario occurs. 
Fig.\ref{ftmonster}a shows the spectral density of $<uv>$ 
of the structure that leads to the collapsing monster. 
It closely agrees to Fig.\ref{ftmonster}b which shows the spectral density 
for smooth initial conditions of Fig.\ref{kdvmonster}d 
related to the continuous system (\ref{dkdv}). 
The similarity to the collapse in twodimensional 
NLS systems suggests that a net flow of the spectral density of $<uv>$ towards 
small wave numbers moves the system towards a saddle point for the 
heteroclinic connection to infinity, but the driving force of this 
process is not yet understood. 
\section{Conclusions}
The fusing of the peaks that emerge from an initial phase instability is 
distinctive for the self focusing in nonintegrable systems. As opposed 
to integrable dispersive nonlinear systems, it leads to the formation 
of coherent structures where peaks of high amplitude emerge from 
a disordered low amplitude background. Small wavelength radiation 
emitted by the fusing peaks leads to an 
 irreversible transfer of energy to small scales. This can be understood 
as homoclinc chaos at the onset of an Arnold diffusion process where the 
trajectory separates from a critical torus to less distinctive parts of 
the energy shell in phase space. \\
This process can be interpreted as the thermalization of energy of an 
ordered initial state. The table \ref{tableap} summarizes these results. 
\begin{table}
\begin{tabular}{|l|l|l|l|}\hline
{\bf Integrals of motion} &
{\bf Initial conditions} &
{\bf Phenomena} &
{\bf Path to maximum}
\\ 
&&&{\bf entropy}
\\
\hline
1. Hamiltonian $\cal{H}$ &
long waves with & 
Formation of coherent &
transfer of $<\cal{N}>$ into
\\ 2. $\cal{M}$ (magnetization) &
low amplitudes& 
structures and low-&
 coherent structures
\\or $\cal{N}$ (particle number)&
low ratio & 
amplitude fluctuations &
optimum ratio  
\\
due to rotational &$<\cal{H}>/<\cal{N}>$ 
&(Figs.\ref{spinsketch}, \ref{sprofile}, 
&of $<\cal{H}>/<\cal{N}>$ 
\\
symmetry 
&&
\ref{smax}a,b, \ref{nlsmax}a)
&in fluctuations, 
\\ (section 2.1.3)
&&&
(Figs.\ref{entropyfig}a, \ref{eqpower}a)
\\ \hline
1. Hamiltonian $\cal{H}$ &
short waves with & 
Destruction of coherent &
transfer of $<\cal{N}>$ from 
\\ 2. $\cal{M}$ or $\cal{N}$ due to&
small amplitudes and& 
structures &
 coherent structures into 
\\rotational symmetry&
high $<\cal{H}>/<\cal{N}>$; & 
(Fig.\ref{smax}c)&
fluctuations; decrease   
\\
(section 2.1.3)
&coherent structures&
&of $<\cal{H}>/<\cal{N}>$ 
\\
&with low 
&&in fluctuations 
\\ 
&$<\cal{H}>/<\cal{N}>$&&
(Figs.\ref{entropyfig}a, \ref{eqpower}b)
\\ \hline
Hamiltonian $\cal{H}$ &
any& 
disordered fluctuations&
production or reduction 
\\
broken rotational &
&(Figs.\ref{nlsmax}c,d)
&
of $<\cal{N}>$  
\\symmetry&
& 
&
optimum ratio    
\\
(section 3.3.4)
&&
&of $<\cal{H}>/<\cal{N}>$ 
\\
&&&in fluctuations
\\ \hline
Hamiltonian $\cal{H}$ &
any& 
quasiperiodic emergence &
no thermalization
\\
infinite number of &
&of coherent structures
& 
\\integrals of motion&
&(Fig.\ref{nlsmax}b) 
&  
\\
(section 3.3.3)
&&& 
\\ \hline
1. energy $\sum uv$ &
$k=2\pi/3$-waves with& 
Formation of coherent &
transfer of $\sum uv$ into
\\ 2. mass $\sum u$, $\sum v$ 
&nonzero energy
 & 
structures and low-&
 coherent structures,
\\(section 2.2.2)
&
& 
amplitude fluctuations &
heteroclinic connection  
\\
& 
&(Fig.\ref{kdvmax}a);
&to infinity
\\
&&'monster' (Fig.\ref{kdvmonster}c)
&(Fig.\ref{uvprofile})
\\ \hline
1. energy $\sum uv$ &
white noise with& 
sudden local growth&
heteroclinic connection  
\\ 2. mass $\sum u$, $\sum v$ 
&zero energy
 & 
of the amplitude&
to infinity
\\(section 2.2.2)
&
& 
(Figs.\ref{kdvmax}b, \ref{kdvmonster}a,b);
&(Fig.\ref{kdvmonster}c)
\\
& 
&'monster' (Fig.\ref{kdvmonster}c)
& 
\\ \hline
\end{tabular}
\label{tableap}
\end{table}
We have discussed the equilibrium thermodynamics 
of a low temperature state that is reached after a long time for a 
generic model, the 
anisotropic Heisenberg spin chain. The system reaches a state where 
most spins contribute to a disordered state near the north pole while 
the coherent structures correspond to a few xenochrysts where the spins 
point to the south pole (first row in the table). 
A similar state is reached in nonintegrable 
NLS type of equations but it is absent in integrable NLS equations 
(fourth row in the table) 
and in systems that do not contain a second integral of motion in addition 
to the energy (third row). 
The latter thermalize in a low amplitude state with no coherent 
structures while integrable systems show continuing quasiperiodic motion. \\
The physical conclusion of this result is that the focusing process 
is driven by the generation of entropy in a state of small 
amplitude waves. 
The link between 
the entropy and the emergence of peaks is the constraint imposed by two 
integrals of motion. Starting from a highly ordered state, 
the system can not simply increase its entropy by reaching its most 
likely state of small amplitude fluctuations since it is restricted 
by a second conservation law. To reach the entropy maximum, 
the system has to allocate the right amount of the second 
conserved quantity to low amplitude fluctuations that absorb 
most of the energy. Whenever there is a surplus of this second 
conserved quantity, 
this maximum can be reached by gathering the surplus of the second conserved 
quantity in the sites of the coherent structures 
(first row of the table). 
The system cannot thermalize completely and no coherent 
structures emerge if there is a lack of the second conserved quantity 
(second row). 
This leads to the phase transition-like dependence of the amount of 
coherent structures on the energy in Fig.7. 
\\
Speaking in terms of equation 
(\ref{power}), an initial distribution of particles $n_k$ will 
not be able to reach the Rayleigh-Jeans distribution while obeying 
both the conservation of energy $\sum n_k\omega_k$ and the particle 
number $\sum n_k$. But the restriction of particle conservation may 
be circumvented by gathering low-energy particles in small domains 
and transferring their energy to the remaining particles to 
increase the overall entropy. In that sense, the formation 
of coherent structures is reminiscent of the condensation 
of droplets in oversaturated steam where the entropy is maximized under 
the restriction of particle conservation. \\
The conserved quantities of the leapfrog-discretized KdV equation 
are insufficient to ensure a thermalization in a well-defined state. 
This  system contains new degrees of freedom since the amplitudes at even 
and at odd times may not be in step. 
The conserved correlation $2\sum u(n)v(n)$ of these fields is 
the difference of the physical 
$\sum (u(n)+v(n))^2/2$ and computational $\sum(u(n)-v(n))^2/2$ energies  
that both can grow in an unbounded fashion 
similar to $\int |\nabla \phi|^2 dV$ 
and $\int |\phi|^4 dV$ in the collapse of nonlinear Schr\"odinger systems. 
An infinite phase space volume is accessible on the shells of 
constant $\sum u(n)v(n)$. 
If $\sum u(n)v(n)$ is finite 
(fifth row of the table), a modulational instability 
of usual Benjamin-Feir type leads to traveling soliton waves which sweep up 
smaller ones as they travel along the circle 
so that a scenario of merging peaks takes place on intermediate time scales. 
Finally, the system 
finds a way to exploit all of the phase space when a long-wave 
instability of the background noise gives rise to the rapidly 
growing 'monster'-solution. The same solution can be initiated by a 
short-wavelength instability for $\sum u(n)v(n)=0$ (sixth row). 
This solution is the systems canonical way 
of exploiting the phase space associated with infinitely high amplitudes. 
\section{Acknowledgement}
B.R. gratefully acknowledges support by a grant of German Academic 
Exchange Service (DAAD). A.C.N gratefully acknowledges support 
from NSF grant DMS 0072803.

\end{document}